\documentclass[twoside,leqno]{article}
\usepackage{ltexpprt}
\usepackage{floatflt,graphicx}
\usepackage{amssymb}
\usepackage{amsfonts,amsmath}
\usepackage{algorithm2e, framed}
\usepackage{multirow}

\def\C1{\hbox{{\rm C}\hskip -5pt\vrule height 7pt width .7pt depth -1pt\hskip 5pt}}
\newenvironment{proof}[1][Proof]{\begin{trivlist}
\item[\hskip \labelsep {\bfseries #1}]}{\end{trivlist}}

\newcommand{\qed}{\nobreak \ifvmode \relax \else
      \ifdim\lastskip<1.5em \hskip-\lastskip
      \hskip1.5em plus0em minus0.5em \fi \nobreak
      \vrule height0.75em width0.5em depth0.25em\fi}

\begin{document}

\title{
Simrank++: Query rewriting through link analysis of the click
graph}
\author{
Ioannis Antonellis
\thanks{Computer Science Dept, Stanford
University. Email: \texttt{antonell@cs.stanford.edu}} \and Hector
Garcia-Molina
\thanks{Computer Science Dept, Stanford University. Email:
\texttt{hector@cs.stanford.edu}} \and Chi-Chao Chang
\thanks{Yahoo! Inc. Email:
\texttt{chichao@yahoo-inc.com}}  }
\date{}
\maketitle

\pagestyle{headings} \pagenumbering{arabic}

\begin{abstract}
We focus on the problem of query rewriting for sponsored search.
We base rewrites on a historical click graph that records the ads
that have been clicked on in response to past user queries. Given
a query $q$, we first consider Simrank~\cite{DBLP:confs/KDD02} as
a way to identify queries similar to $q$, i.e., queries whose ads
a user may be interested in. We argue that Simrank fails to
properly identify query similarities in our application, and we
present two enhanced versions of Simrank: one that exploits
weights on click graph edges and another that exploits
``evidence.'' We experimentally evaluate our new schemes against
Simrank, using actual click graphs and queries form Yahoo!, and
using a variety of metrics. Our results show that the enhanced
methods can yield more and better query rewrites.
\end{abstract}

\section{Introduction}
\label{sec:introduction}

In sponsored search, paid advertisements (ads) relevant to a
user's query are shown above or along-side traditional web search
results. The placement of these ads is in general related to a
ranking score which is a function of the semantic relevance to the
query and the advertiser's bid.
\begin{figure}[htbp]
 \begin{center}
\includegraphics[scale=.20,angle=0]{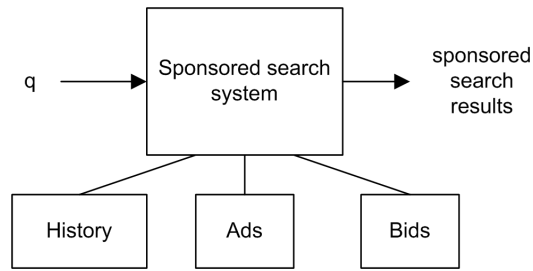}
\caption{General sponsored search system architecture.}
\label{fig:system1}
\end{center}
\end{figure}

Ideally, a sponsored search system would appear as in Figure
\ref{fig:system1}. The system has access to a database of
available ads and a set of bids. Conceptually, each bid consists
of a query $q$, an ad $\alpha$, and a price $p$. With such a bid,
the bidder offers to pay if the ad $\alpha$ is both displayed and
clicked when a user issues query $q$. For many queries, there are
not enough direct bids, so the sponsored search system attempts to
find other ads that may be of interest to the user who submitted
the query. Even though there is no direct bid, if the user clicks
on one of these ads, the search engine will make some money (and
the advertiser will receive a customer). The challenge is then to
find ads related to incoming queries that may yield user click
throughs.

\begin{figure}[htbp]
 \begin{center}
\includegraphics[scale=.20,angle=0]{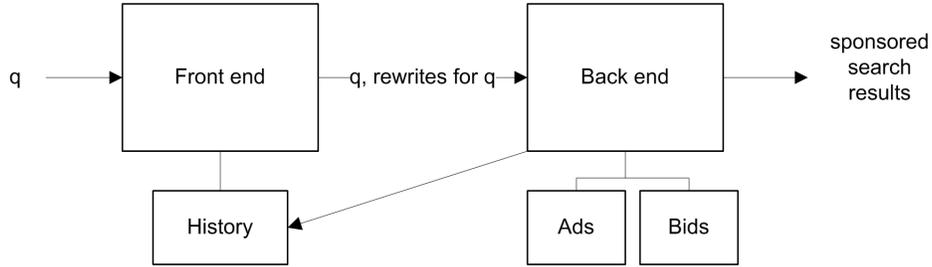}
\caption{A common sponsored search system architecture.}
\label{fig:system2}
\end{center}
\end{figure}
For a variety of practical and historical reasons, the sponsored
search system is often split into two components, as shown in
Figure \ref{fig:system2}. A front-end takes an input query $q$ and
produces a list of {\em re-writes}, i.e., of other queries that
are ``similar'' to $q$. For example, for query ``camera,'' the
queries ``digital camera'' and ``photography'' may be useful
because the user may also be interested in ads for those related
queries. The query ``battery'' may also be useful because users
that want a camera may also be in the market for a spare battery.
The query and its rewrites are then considered by the back-end,
which displays ads that have bids for the query or its rewrites.
The split approach reduces the complexity of the back-end, which
has to deal with rapidly changing bids. The work of finding
relevant ads, indirectly through related queries, is off-loaded to
the front-end.

At the front-end, queries can be rewritten using a variety of
techniques (reviewed in our Related Work section) developed for
document search. However, these techniques often do not generate
enough useful rewrites. Part of the problem is that in our case
``documents'' (the ads) have little text, and queries are very
short, so there is less information to work with, as compared with
larger documents. Another part of the problem is that there are
relatively few queries in the bid database, so even if we found
all the textually related ones, we may not have enough. Thus, it
is important to generate additional rewrites, using other
techniques.

In this paper we focus on query rewrites based on the recent
history of ads displayed and clicked on. The back-end generates a
historical {\em click graph} that records the clicks that were
generated by ads when a user inputs a given query. The click graph
is a weighted bi-partite graph, with queries on one side and ads
on the other (details in Section \ref{sec:notation}). The schemes
we present analyze the connections in the click graph to identify
rewrites that may be useful. Our techniques identify not only
queries that are directly connected by an ad (e.g., users that
submit either ``mp3'' or ``i-tunes'' click on ad an for ``iPod.'')
but also queries that are more indirectly related (Section
\ref{sec:similar_queries}). Our techniques are based on the notion
of {\em SimRank} \cite{DBLP:confs/KDD02}, which can compute query
similarity based on the connections in a bi-partite click-graph.
However, in our case we need to extend SimRank to take into
account the specifics of our sponsored search application.

Briefly, the contributions of this paper are as follows.
\begin{itemize}

\item We present a framework for query rewriting in a sponsored
search environment.

\item We identify cases where SimRank fails to transfer correctly
the relationships between queries and ads into similarity scores.

\item We present two SimRank extensions: one that takes into
account the weights of the edges in the click graph, and another
that takes into account the ``evidence'' supporting the similarity
between queries.

\item We experimentally evaluate these query rewriting techniques,
using an actual click graph from Yahoo!, and a set of queries
extracted from Yahoo! logs. We evaluate the resulting rewrites
using several metrics. One of the comparisons we perform involves
manual evaluation of query-rewrite pairs by members of Yahoo!'s
Editorial Evaluation Team. Our results show that we can
significantly increase the number of useful rewrites over those
produced by SimRank and by another basic technique.

\end{itemize}

\subsection{Related Work}
\label{sec:related_work}

The query rewriting problem has been extensively studied in terms
of traditional web search. In traditional web search, query
rewriting techniques are used for recommending more useful queries
to the user and for improving the quality of search results by
incorporating users' actions in the results' ranking of future
searches. Given a query and a search engine's results on this, the
indication that a user clicked on some results can be interpreted
as a vote that these specific results are matching the user's
needs and thus are more relevant to the query. This information
can then be used for improving the search results on future
queries. Existing query rewriting techniques for traditional web
search, include relevance feedback and pseudo-relevance feedback,
query term deletion \cite{DBLP:confs/JF03}, substituting query
terms with related terms from retrieved documents
\cite{DBLP:confs/TCCIMK04}, dimensionality reduction such as
Latent Semantic Indexing (LSI) \cite{DDLFH1990}, machine learning
techniques \cite{DBLP:confs/ZHRJ07, WNZATIS2002,
DBLP:confs/BB2000} and techniques based on the analysis of the
click graph \cite{DBLP:confs/CS2007}.

Pseudo-relevance feedback techniques involve submitting a query
for an initial retrieval, processing the resulting documents,
modifying the query by expanding it with additional terms from the
documents retrieved and then performing an additional retrieval on
the modified query. However, pseudo-relevance feedback requires
that the initial query retrieval procedure returns some results,
something that is not always the case in sponsored search, as
described before. In addition, pseudo-relevance has many
limitations in effectiveness \cite{DBLP:confs/RSIGIR03}. It may
lead to query drift, as unrelated terms might be added to the
query and is also computationally expensive. Query relaxation or
deleting query terms leads to a loss of specificity from the
original query.

In LSI, a collection of queries is represented by a terms queries
matrix where each column corresponds to the vector space
representation of a query. The column space of that matrix is
approximated by a space of much smaller dimension that is obtained
from the leading singular vectors of the matrix and then
similarity scores between different queries can be computed. LSI
is frequently found to be very effective even though the analysis
of its success is not as straightforward
\cite{DBLP:confs/PTRVPODS98}. The computational kernel in LSI is
the singular value decomposition (SVD). This provides the
mechanism for projecting both the queries  on a lower-dimensional
space spanned by the leading left singular vectors. In addition to
performing dimensionality reduction, LSI captures hidden semantic
structure in the data and resolves problems caused by synonymy and
polysemy in the terms used. However, a well known difficulty with
LSI is the high cost of the SVD for the large, sparse matrices
appearing in practice.

\section{Problem Definition} \label{sec:notation}

Let $\mathcal{Q}$ denote a set of $n$ queries and $\mathcal{A}$
denote a set of $m$ ads. A click graph for a specific time period
is an undirected, weighted, bipartite graph $G = (\mathcal{Q},
\mathcal{A}, E)$ where $E$ is a set of edges that connect queries
with ads. $G$ has an edge $(q,\alpha)$ if at least one user that
issued the query $q$ during the time period also clicked on the ad
$\alpha$. Each edge $(q,\alpha)$ has three weights associated with
it. The first one is the number of times that $\alpha$ has been
displayed as a result for $q$ and is called the impressions of
$\alpha$ given $q$. The second weight is the number of clicks that
$\alpha$ received as a result of being displayed for they query
$q$.  This second weight is less than or equal to the first
weight. The number of clicks divided by the number of impressions
gives us the likelihood that a displayed ad will be clicked on.
However, to be more accurate, this ratio needs to be adjusted to
take into account the position where the ad was displayed. That
is, an ad $\alpha$ placed near the top of the sponsored results is
more likely to be clicked on, regardless of how good an ad it is
for query $q$. Thus, the third weight associated with an edge
$(q,\alpha)$ is the {\em expected click rate}, an adjusted clicks
over impressions rate. The expected click rate is computed by the
back-end (Figure \ref{fig:system2}), and we do not discuss the
details here.

Finally, for a node $v$ in a graph, we denote by $E(v)$ the set of
neighbors of $v$. We also define $N(v) = |E(v)|$ that is $N(v)$
denotes the number of $v$'s total neighbors.

As discussed in the introduction, our goal is to find queries that
are {\em similar}, in the sense that the ads clicked on for one
query are likely to be clicked on when displayed for a user that
entered the second query. We will predict similarity based on the
information in the click graph: The intuition is that if an ad
received clicks when displayed for both queries $q_1$ and $q_2$,
then the queries are similar. Furthermore, if $q_2$ is related to
$q_3$ in the same way but through some other ad, then $q_1$ and
$q_3$ are also similar, although possibly to a lesser degree. We
discuss our notion of similarity more in the following section.

Note that if the click graph does {\em not} contain an ad $\alpha$
that received clicks when $q_1$ and $q_2$ were issued, then we
{\em cannot} infer that $q_1$ and $q_2$ are {\em not} similar. The
queries could very well be similar (in our sense), but while the
click-graph was collected, the back-end did not display ads that
would have shown this similarity. (Perhaps there were no bids for
those ads at the time.) As we will see later, even without the
common ad $\alpha$, we may still be able to discover the
similarity of $q_1$ and $q_2$ through other similarity
relationships in the click-graph.

Also note that in this paper we are not addressing problems of
click or ad fraud. Fraud is a serious problem, where organizations
or individuals generate clicks or place ads with the intent of
defrauding or misleading the advertiser and/or the search engine.
Query rewriting strategies may need to be adjusted to protect from
fraud, but we do not consider such issues here.

Finally, notice that our query rewriting problem is a type of
collaborative filtering (CF) problem. We can view the queries as
``users'' who are recommending ``ads'' by clicking on them. When
we identify similar queries, we are finding queries that have
similar recommendations, just like in CF, where one finds users
that have similar tastes. In our setting, we are only trying to
find similar queries (users), and not actually predicting
recommended ads. Furthermore, as we will see, we are tuning our
similarity metrics so they work well for sponsored search, as
opposed to generic recommendations.

\section{Similar queries} \label{sec:similar_queries}
\begin{figure}[htbp]
 \begin{center}
\includegraphics[scale=.25,angle=0]{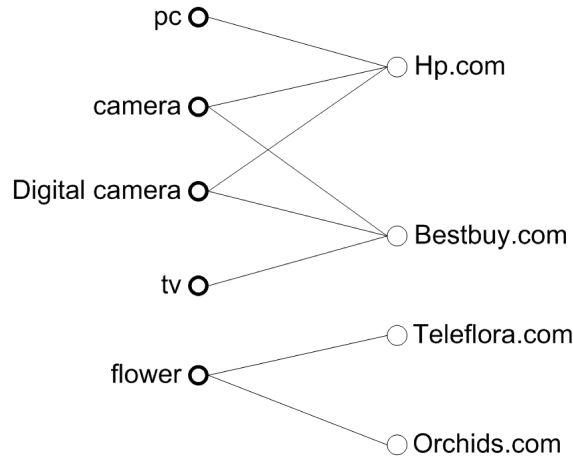}
\caption{Sample unweighted click graph. An edge indicates the
existence of at least one click from a query to an ad.}
\label{example_click_graph}
\end{center}
\end{figure}

In this section we discuss the notion of query similarity that we
are interested in. As we mentioned earlier, we will be saying that
two queries are similar if they tend to make the search engine
users to click on the same ads. Let us illustrate this with an
example. Figure \ref{example_click_graph} shows a small click
graph; for simplicity we have removed the weights from the edges
and thus an edge indicates the existence of at least one click
from a query to an ad. In this graph, the queries ``pc'' and
``camera'' are connected through a common ad and thus can be
considered similar. Notice that this notion of similarity is not
related to the actual similarity of the concepts described by the
query terms. Now, we can observe that the queries ``camera'' and
``digital camera'' are connected through two common ads and thus
can be considered similar. In contrast, queries ``pc'' and ``tv''
are not connected through any ad. However, both ``pc'' and ``tv''
are connected through an ad with the queries ``digital camera''
and ``camera'' which we already saw that are similar. Thus, we
have a small amount of evidence that ``pc'' and ``tv'' are somehow
similar, because they are both similar with queries that bring
clicks to the same ads. In that case we will be saying that ``pc''
and ``tv'' are one hop away from queries that have a common ad.
There might actually be cases where two queries will be two or
more hops away from queries that bring clicks to the same ad.
Finally, let us consider the queries ``tv'' and ``flower''. There
is no path in the click graph that connects these two queries and
thus we conclude that these queries are not similar.

Thus, a naive way to measure the similarity of a pair of queries
would be to count the number of common ads that they are connected
to. Table \ref{naive_similarity_scores} presents the resulting
similarity scores for our sample click graph. As we can see there,
``pc'' has a similarity score 1 both with ``camera'' and ``digital
camera'' but no similarity with ``tv'' and ``flower''. However,
``camera'' has a similarity score 2 with ``digital camera'' which
indicates a stronger similarity. Also, ``tv'' has similarity 0
both with ``pc'' and ``flower''. Notice also that flower has
similarity 0 with all the other queries. It is obvious that this
naive technique cannot capture the similarity between  ``pc'' and
``tv'' (as it does not look at the whole graph structure) and
determines that their similarity is zero. In the following section
we will see how we can compute similarity scores that take into
account all the interactions appearing in the graph.
\begin{table}
\begin{center}
\caption{Query-query similarity scores for the sample click graph
of Figure \ref{example_click_graph}. Scores have been computed by
counting the common ads between the queries}
\begin{tabular} {|r||c|c|c|c|c|}
\hline  & pc & camera & digital camera&
 tv & flower\\
 \hline
 \hline
 pc & -& 1 & 1 & 0 & 0\\
 \hline
 camera & 1& - & 2 & 1 & 0\\
 \hline
 digital camera & 1& 2 & -  & 1 & 0\\
 \hline
 tv &0 & 1 & 1 & - & 0\\
 \hline
 flower & 0& 0 & 0 & 0 & -\\
 \hline
\end{tabular}
\label{naive_similarity_scores}
\end{center}
\end{table}

\section{Simrank-based query similarity}
\label{sec:simrank_based_similarities} Simrank
\cite{DBLP:confs/KDD02} is a method for computing object
similarities, applicable in any domain with object-to-object
relationships, that measures similarity of the structural context
in which objects occur, based on their relationships with other
objects. Specifically, in the case where there are two types of
objects, bipartite Simrank is an iterative technique to compute
the similarity score for each pair of objects of the same type.
Bipartite Simrank is based on the underlying idea that two objects
of one type are similar if they are related to similar objects of
the second type. In our case, we can consider the queries as one
type of objects and the ads as the other and use bipartite Simrank
to compute similarity scores for each query-query pair.

Let $s(q, q')$ denote the similarity between queries $q$ and $q'$,
and let $s(\alpha, \alpha')$ denote the similarity between ads
$\alpha$ and $\alpha'$. For $q \neq q'$, we write the equation:
\begin{equation}
s(q, q') = \frac{C_1}{N(q)N(q')} \sum_{i \in E(q)} \sum_{j \in
E(q')} s(i,j) \label{eq:query_similarities}
\end{equation}
where $C_1$ is a constant between 0 and 1. For $\alpha \neq
\alpha'$, we write:
\begin{equation}
s(\alpha, \alpha') = \frac{C_2}{N(\alpha) N(\alpha')} \sum_{i \in
E(\alpha)} \sum_{j \in E(\alpha')} s\left(i, j\right)
\label{eq:ad_similarities}
\end{equation}
where again $C_2$ is a constant between 0 and 1.

If $q = q'$, we define $s(q, q') = 1$ and analogously if $\alpha =
\alpha'$ we define $s(\alpha, \alpha') = 1$. Neglecting $C_1$ and
$C_2$, equation~\ref{eq:query_similarities} says that the
similarity between queries $q$ and $q'$ is the average similarity
between the ads that were clicked on for $q$ and $q'$. Similarly,
equation~\ref{eq:ad_similarities} says that the similarity between
ads $\alpha$ and $\alpha'$ is the average similarity between the
queries that triggered clicks on $\alpha$ and $\alpha'$.

In the SimRank paper~\cite{DBLP:confs/KDD02}, it is shown that a
simultaneous solution $s(*, *) \in [0,1]$ to the above equations
always exists and is unique. Also notice that the SimRank scores
are symmetric, i.e. $s(q, q') = s(q', q)$.

In order to understand the role of the $C_1, C_2$ constants, let
us consider a simple scenario were two ads $\alpha$ and $\alpha'$
were clicked on for a query $q$ (which means that edges from $q$
towards $\alpha$ and $\alpha'$ exist), so we can conclude some
similarity between $\alpha$ and $\alpha'$. The similarity of $q$
with itself is 1, but we probably don't want to conclude that
$s(\alpha, \alpha') = s(q, q) = 1$. Rather, we let $s(\alpha,
\alpha') = C_2 \cdot s(q,q)$, meaning that we are less confident
about the similarity between $\alpha$ and $\alpha'$ than we are
between $q$ and itself.

Let us look now at the similarity scores that Simrank computes for
our simple click graph of Figure \ref{example_click_graph}. Table
\ref{simrank_similarity_scores} presents the similarity scores
between all query pairs. If we compare these similarity scores
with the ones in Table \ref{naive_similarity_scores}, we can make
the following observations. Firstly, ``camera'' and ``digital
camera'' have now the same similarity score with all other queries
except for ``flower''. Secondly, ``tv'' has similarity $0.437$
with ``pc'', $0.619$ with ``camera'' and ``digital camera'' and
zero with ``flower''. Notice that Simrank takes into account the
whole graph structure and thus correctly produces a nonzero
similarity score for the pair ``tv'' - ``pc''. Also notice that
``camera'' has two common ads with ``digital camera'' and only one
common ad with ``tv''. However, Simrank does not produce different
similarity scores for the ``camera''-``digital camera'' and
``camera''-``tv'' pairs. We will come back to this issue in detail
in Section \ref{sec:complete_bipartite_graphs}.
\begin{table}
\begin{center}
\caption{Query-query similarity scores for the sample click graph
of Figure \ref{example_click_graph}. Scores have been computed by
Simrank with $C_1 = C_2 = 0.8$}
\begin{tabular} {|r||c|c|c|c|c|}
\hline  & pc & camera & digital camera&
 tv & flower\\
 \hline
 \hline
 pc & - & 0.619 & 0.619 & 0.437 & 0\\
 \hline
 camera & 0.619& - & 0.619 & 0.619 & 0\\
 \hline
 digital camera & 0.619& 0.619 & -  & 0.619 & 0\\
 \hline
 tv &0.437 & 0.619 & 0.619 & - & 0\\
 \hline
 flower & 0& 0 & 0 & 0 & - \\
 \hline
\end{tabular}
\label{simrank_similarity_scores}
\end{center}
\end{table}
\section{Random walks behind Simrank}
\label{sec:random_walks_behind_simrank} The intuition behind the
similarity scores that Simrank defines is based on a ``random
surfers'' model. According to this, a Simrank score
$\textrm{sim}(a, b)$ measures how soon two random surfers are
expected to meet at the same node if they started at nodes $a$,
$b$ and randomly walked the graph. The transition probabilities of
this random walk are uniform, which means that (assuming $C_1 =
C_2 = 1$) if $a$ has $n$ out-neighbors, with the same probability
$1/n$ the random surfer will move to one of these out-neighbors.

The decay factors $C_1, C_2$ allow for self-transitions.
Self-transitions correspond to transitions from a node to itself.
$C_1$ affects the self-transition probabilities of one of the
graph's node sets while $C_2$ affects the self-transition
probabilities of the other node set. Given that $C_1 < 1$, with
probability $1 - C_1$ a random surfer will remain in the same node
and with probability $C_1/n$ he will move to one of the $n$
out-neighbors of the node.

\section{Simrank in complete bipartite graphs}
\label{sec:complete_bipartite_graphs} 
\begin{figure}[htbp]
 \begin{center}
\includegraphics[scale=.20,angle=0]{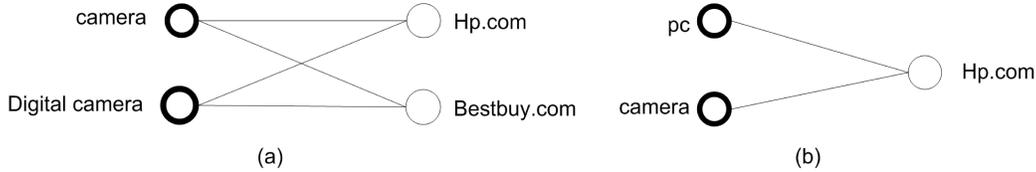}
\caption{Sample complete bipartite graphs ($K_{2,2}$ and
$K_{1,2}$) extracted from a click graph. }
\label{complete_bipartite_graphs_examples}
\end{center}
\end{figure}

Some simple bipartite graphs
that often appear as subgraphs of a click graph are the complete
bipartite graphs. A complete bipartite graph is a special kind of
bipartite graph where every vertex of the first node set is
connected to every vertex of the second nodes set. In the click
graph of Figure \ref{example_click_graph}, the subgraphs
consisting of the nodes ``flower'', ``Teleflora.com'',
``orchids.com'' and ``camera'', ``digital camera'', ``hp.com'',
``bestbuy.com'' are two examples of complete bipartite subgraphs.
Formally, a complete bipartite graph $G = (V_1,V_2, E)$ is a
bipartite graph such that for any two vertices $v_1 \in V_1$ and
$v_2 \in V_2$, $(v_1, v_2)$ is an edge in $E$. The complete
bipartite graph with partitions of size $|V_1| = m$ and $|V_2| =
n$, is denoted $K_{m,n}$. Figure
\ref{complete_bipartite_graphs_examples}(a) shows a $K_{2,2}$
graph from a click graph and Figure
\ref{complete_bipartite_graphs_examples}(b) shows a $K_{1,2}$
click graph.

Let us look at the similarity scores that Simrank computes for the
pairs ``camera'' - ``digital camera'' and ``pc'' - ``camera'' from
the graphs of Figure \ref{complete_bipartite_graphs_examples}.
Table \ref{simrank_similarity_scores_bipartite} tabulates these
scores for the first 7 iterations. As we can see sim(``camera'',
``digital camera'') is always less than sim(``pc'', ``camera'')
although we observe that sim(``camera'', ``digital camera'')
increases as we include more iterations. In fact, we can prove
that sim(``camera'', ``digital camera'') becomes eventually equal
to sim(``pc'', ``camera'') as we include more iterations.
\begin{table}
\begin{center}
\caption{Query-query similarity scores for the sample click graphs
of Figure \ref{complete_bipartite_graphs_examples}. Scores have
been computed by Simrank with $C_1 = C_2 = 0.8$}
\begin{tabular} {|r||r|r|}
\hline
Iteration & sim(``camera'', ``digital camera'') & sim(``pc'', ``camera'')\\
 \hline
1 & 0.4 & 0.8\\
 \hline
2 & 0.56 & 0.8\\
 \hline
3 & 0.624 & 0.8\\
 \hline
4 & 0.6496 & 0.8\\
 \hline
5 & 0.65984 & 0.8\\
 \hline
6 & 0.663936 & 0.8\\
 \hline
7 & 0.6655744 & 0.8\\
 \hline
\end{tabular}
\label{simrank_similarity_scores_bipartite}
\end{center}
\end{table}
We can actually  prove the following two Theorems for the
similarity scores that Simrank computes in complete bipartite
graphs (refer to Appendix \ref{sec:appendix1} for the proofs).
\begin{theorem}
Consider the two complete bipartite graphs $G = K_{1,2}$ and $G' =
K_{2,2}$ with nodes sets $V_1 = \{a\}, V_2 = \{A, B\}$ and $V'_1 =
\{b, c\}$ and $V'_2 = \{C,D\}$ correspondingly. Let
$\textrm{sim}^{(k)}(A, B)$ and $\textrm{sim}^{(k)}(C, D)$ denote
the similarity scores that bipartite Simrank computes for the node
pairs $(A,B)$ and $(C, D)$ after $k$ iterations. Then,
$\textrm{sim}^{(k)}(A, B) \geq \textrm{sim}^{(k)}(C,D),\ \forall \
k>0$. \qed
\end{theorem}
\begin{theorem}
Consider the two complete bipartite graphs $G = K_{m,2}$ and $G' =
K_{n,2}$ with $m < n$ and nodes sets $V_1, V_2 = \{A, B\}$ and
$V'_1, V'_2 = \{C,D\}$ correspondingly. Let $\textrm{sim}^{(k)}(A,
B)$ and $\textrm{sim}^{(k)}(C, D)$ denote the similarity scores
that bipartite Simrank computes for the node pairs $(A,B)$ and
$(C, D)$ after $k$ iterations. Then,
\begin{itemize}
\item[(i)] $\textrm{sim}^{(k)}(A, B) > \textrm{sim}^{(k)}(C,D),\
\forall \ k>0$, and
\item[(ii)]$\lim_{k \rightarrow \infty} \
\textrm{sim}^{(k)}(A, B) = \lim_{k \rightarrow \infty}\
\textrm{sim}^{(k)}(C,D)$ if and only if $C_1 = C_2 = 1$, where
$C_1,\ C_2$ are the decay factors of the bipartite Simrank
equations.\qed
\end{itemize}
\end{theorem}
These Theorems provide us two pieces of evidence that Simrank
scores are not intuitively correct in complete bipartite graphs.
First, as in practice Simrank computations are limited to a small
number of iterations, we would reach the conclusion that the pair
``pc''-``camera'' is more similar than the pair ``camera'' -
``digital camera'' which is obviously not correct. Second, even if
we had the luxury to run Simrank until it converges, we would
reach the conclusion that the similarity scores of the two pairs
are the same. However, the fact that there are two advertisers
that are connected with the queries ``camera'' and ``digital
camera'' (versus the one that connects ``pc'' with ``camera'') is
an indication that their similarity is stronger. We will try to
fix such cases by introducing the notion of ``evidence of
similarity'' in the following section.

\section{Revising Simrank}
\label{sec:fixing_simrank} Consider a bipartite graph $G = (V_1,
V_2, E)$ and two nodes $a, b \in V_1$. We will denote as
$\textrm{evidence}(a, b)$ the evidence existing in $G$ that the
nodes $a, b$ are similar. The definition of $\textrm{evidence}(a,
b)$ we use is shown on Equation \ref{evidence_score1}.
\begin{equation}
\textrm{evidence}(a, b) = \sum_{i=1}^{|E(a) \bigcap E(b)|}
\frac{1}{2^i} \label{evidence_score1}
\end{equation}
The intuition behind choosing such a function is as follows. We
want the evidence score evidence(a,b) to be an increasing function
of the  common neighbors between a and b. In addition we want the
evidence scores to get closer to one as the common neighbors
increase. Thus, another reasonable choice would be the following:
\begin{equation}
\textrm{evidence}(a, b) = \left( 1 - e^{-|E(a) \bigcap
E(b)|}\right)\label{evidence_score2}
\end{equation}
In our experiments we used the first definition although
preliminary results with both formulas did not show substantial
differences.

 We can now incorporate the evidence metric into the Simrank
equations. We modify the equations \ref{eq:query_similarities} and
\ref{eq:ad_similarities} as follows:

For $q \neq q'$, we write the equation:
\begin{eqnarray}
s_\textrm{evidence}(q, q') &=& \textrm{evidence}(q, q') \cdot s(q,
q')
 \label{eq:query_similarities_fixed}
\end{eqnarray}
where $s(q,q')$ is the Simrank similarity between $q$ and $q'$.
For $\alpha \neq \alpha'$, we write:
\begin{eqnarray}
s_\textrm{evidence}(\alpha, \alpha') &= &\textrm{evidence}(\alpha,
\alpha')\cdot s(\alpha, \alpha')
 \label{eq:ad_similarities_fixed}
\end{eqnarray}
where again $s(\alpha, \alpha')$ is the Simrank similarity between
$\alpha$ and $\alpha'$.

Notice, that we could use $k$ only iterations to compute the
Simrank similarity scores and then multiply them by the evidence
scores to come up with evidence-based similarities after $k$
iterations. We will be loosely referring to these scores as
evidence-based similarity scores after $k$ iterations and we will
be denoting them by $s_\textrm{evidence}^{(k)}(q,q')$.

Let us see now what the new Simrank equations compute for our
sample click graphs.
Table~\ref{table:evidence_simrank_similarity_scores_bipartite}
tabulates these scores. As we can see sim(``camera'', ``digital
camera'') is greater than sim(``pc'', ``camera'') after the first
iteration.
\begin{table}
\begin{center}
\caption{Query-query similarity scores for the sample click graphs
of Figure \ref{complete_bipartite_graphs_examples}. Scores have
been computed by the evidence-based Simrank with $C_1 = C_2 =
0.8$}
\begin{tabular} {|r||r|r|}
\hline
Iteration & sim(``camera'', ``digital camera'') & sim(``pc'', ``camera'')\\
 \hline
1 & 0.3 & 0.4\\
 \hline
2 & 0.42 & 0.4\\
 \hline
3 & 0.468 & 0.4\\
 \hline
4 & 0.4872 & 0.4\\
 \hline
5 & 0.49488 & 0.4\\
 \hline
6 & 0.497952 & 0.4\\
 \hline
7 & 0.4991808 & 0.4\\
 \hline
\end{tabular}
\label{table:evidence_simrank_similarity_scores_bipartite}
\end{center}
\end{table}
We can actually  prove the following Theorem for the similarity
scores that evidence-based Simrank computes in complete bipartite
graphs (refer to Appendix \ref{sec:appendix2} for the proof).
\begin{theorem}
Consider the two complete bipartite graphs $G = K_{m,2}$ and $G' =
K_{n,2}$ with $m < n$ and nodes sets $V_1, V_2 = \{A, B\}$ and
$V'_1, V'_2 = \{C,D\}$ correspondingly. Let $\textrm{sim}^{(k)}(A,
B)$ and $\textrm{sim}^{(k)}(C, D)$ denote the similarity scores
that bipartite evidence-based Simrank computes for the node pairs
$(A,B)$ and $(C, D)$ after $k$ iterations and let $C_1, C_2 >
\frac{1}{2}$, where $C_1,\ C_2$ are the decay factors of the
bipartite Simrank equations. Then,
\begin{itemize}
\item[(i)] $\textrm{sim}^{(k)}(A, B) < \textrm{sim}^{(k)}(C,D),\
\forall \ k>1$, and \item[(ii)]$\lim_{k \rightarrow \infty} \
\textrm{sim}^{(k)}(A, B) < \lim_{k \rightarrow \infty}\
\textrm{sim}^{(k)}(C,D)$. \qed
\end{itemize}
\end{theorem}
This Theorem indicates that the evidence-based Simrank scores in
complete bipartite graphs will be consistent with the intuition of
query similarity (as we discussed it in Section
\ref{sec:similar_queries})  even if we effectively limit the
number of iterations we perform.

\section{Weighted Simrank} \label{sec:weighted_simrank}
In the previous sections we ignored the information contained in
the edges of a click graph and we tried to derive similarity
scores for query pairs by just using the click graph's structure.
In this section, we focus on weighted click graphs. We explore
ways to derive query-query similarity scores that (i) are
consistent with the graph's weights and (ii) utilize the edge
weights in the computation of similarity scores.

\subsection{Consistent similarity scores}
\begin{figure}[htbp]
 \begin{center}
\includegraphics[scale=.23,angle=0]{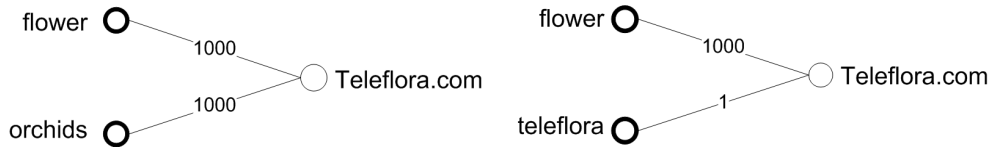}
 \caption{Sample weighted click graphs}
\label{fig:example_weighted_click_graph}
\end{center}
\end{figure}

\begin{figure}[htbp]
 \begin{center}
\includegraphics[scale=.23,angle=0]{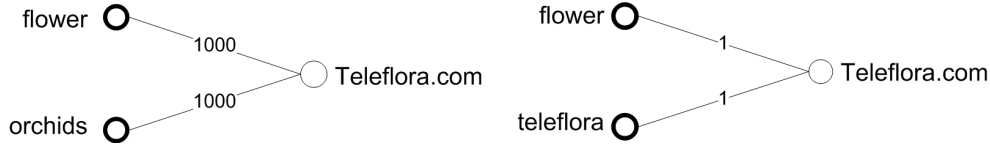}
 \caption{Sample weighted click graphs}
\label{fig:example_weighted_click_graph2}
\end{center}
\end{figure}

We illustrate the notion of consistency between similarity scores
and the graph's weights with the following two examples. Firstly,
consider the two weighted click graphs in  Figure
\ref{fig:example_weighted_click_graph}. Apparently the queries
``flower''-``orchids'' of the left graph are  more ``similar''
than the queries ``flower''-``teleflora'' of the right graph. This
is true because, although both pairs bring clicks to the same ad,
the queries of the first pair bring equally the same amount of
clicks whereas in the second pair the number of clicks each query
brings differ a lot. If we now try to use Simrank or even the
evidence-based Simrank to compute similarity scores for these two
pairs we will see that it will output the exact same similarity
scores for both pairs. It is thus obvious that Simrank scores are
not consistent with the the weights on the graph.
Now, consider the two graphs of Figure
\ref{fig:example_weighted_click_graph2}. Apparently the similarity
scores are no longer affected by the previous notion of
consistency as in both graphs the spread of values of the right
node is the same. However, it is also obvious that now the queries
``flower-orchids'' are more similar than the queries
``flower-teleflora'' since there are more clicks that connect the
first pair with an ad. Again, Simrank or evidence-based Simrank
will output the exact same similarity scores for both pairs.

In general, we define the notion of consistency as follows:
\begin{definition}[Consistent similarity scores]
Consider a weighted bipartite graph $G = (V_1, V_2, E)$. Consider
also two nodes $v_1, v_2 \in V_2$ and four nodes $i_1, j_1, i_2,
j_2 \in V_1$. We now define the sets $W(v_1) = \{w(i_1, v_1),
w(j_1, v_1)\}$ and $W(v_2) = \{w(i_2,v_2), w(j_2, v_2)\}$ and let
$\textrm{variance}(v_1)$ ($\textrm{variance}(v_2)$) denote a
measure of $W(v_1)$'s ($W(v_2)$'s) variance respectively. We will
be saying that a set of similarity scores $\textrm{sim}(i,j)
\forall i,j\in V_1$ is consistent with the graph's weights if and
only if $\forall i_1, j_1, i_2, j_2 \in V_1$ and $\forall v_1,v_2
\in V_2$ such that $\exists (i_1, v_1), (j_1, v_1), (i_2, v_2),
(j_2, v_2) \in E$ both of the following are true:
\begin{itemize}
\item[(i)] If $\textrm{variance}(v_1) = \textrm{variance}(v_2)$
and $w(i_1, v_1) > w(i_2, v_2))$ then $\textrm{sim}(i_1, j_1) >
\textrm{sim}(i_2, j_2)$

\item[(ii)] If $\textrm{variance}(v_1) < \textrm{variance}(v_2)$
and $w(i_1, v_1) > w(i_2, v_2))$ then $\textrm{sim}(i_1, j_1) >
\textrm{sim}(i_2, j_2)$
\end{itemize}
\end{definition}

\subsection{Revising Simrank}
We can now modify the underlying random walk model of Simrank.
Again we use the evidence scores as defined in Section
\ref{sec:fixing_simrank}, but now we will perform a different
random walk. Remember that Simrank's random surfers model implies
that a Simrank score sim($a$, $b$) for two nodes $a$, $b$ measures
how soon two random surfers are expected to meet at the same node
if they started at nodes $a$, $b$ and randomly walked the graph.
In order to impose the consistency rules in the similarity scores
we perform a new random walk where its transition probabilities
$p(\alpha, i),\ \forall \alpha \in V_1, i \in E(\alpha)$ are
defined as follows:
\begin{eqnarray*}
p(\alpha, i) = \textrm{spread}(i) \cdot
\textrm{normalized\_weight}(\alpha,
i), \forall i \in E(\alpha), \textrm{ and}\\
p(\alpha,\alpha) = 1 - \sum_{i \in E(\alpha)} p(\alpha, i)
\end{eqnarray*}
where:
\begin{eqnarray*}
\textrm{spread}(i) = e^{-\textrm{variance}(i)}, \textrm{ and}\\
\textrm{normalized\_weight}(\alpha, i) = \frac{w(\alpha,
i)}{\sum_{j \in E(\alpha)} w(\alpha, j)}
\end{eqnarray*}
Notice how the new transition probability $p(\alpha, i)$ between
two nodes $\alpha \in V_1, i \in V_2$  utilizes both the
$spread(i)$ value and the $w(\alpha, i)$ value in order to satisfy
the consistency rules. Actually, we can prove the following
Theorem that ensures us that weighted Simrank produces consistent
similarity scores.
\begin{theorem}
Consider a weighted bipartite graph $G = (V_1, V_2, E)$ and let
$w(e)$ denote the weight associated with an edge $e \in E$. Let
also sim($i,j$) denote the similarity score that weighted Simrank
computes for two nodes $i,j \in V_1$. Then, $\forall i,j \in V_1$,
sim$(i,j)$ is consistent with the graph's weights.\qed
\end{theorem}
The actual similarity scores that weighted Simrank gives after
applying the modified random walk are:
\begin{eqnarray*}
s_{\textrm{weighted}} (q, q')& =& \textrm{evidence}(q, q') \cdot
C_1 \sum_{i \in E(q)}\sum_{j \in E(q')} W(q, i) W(q', j)
s_{\textrm{weighted}}
(i, j)\\
s_{\textrm{weighted}} (\alpha, \alpha') &=&
\textrm{evidence}(\alpha, \alpha') \cdot C_2 \sum_{i \in
E(\alpha)}\sum_{j \in E(\alpha')} W(\alpha, i) W(\alpha', j)
s_{\textrm{weighted}}
(i, j)\\
\end{eqnarray*}
where the factors $W(q, i)$ and $W(a, i)$ are defined as follows:
\begin{eqnarray*}
W(q, i) &=&  \textrm{spread}(i) \cdot
\textrm{normalized\_weight}(q, i) = e^{-\textrm{variance}(i)}
\frac{w(q, i)}{\sum_{j \in E(q)} w(q, j)},
\textrm{ and}\\
W(\alpha, i ) &=& \textrm{spread}(i) \cdot
\textrm{normalized\_weight}(\alpha, i) = e^{-\textrm{variance}(i)}
\frac{w(\alpha, i)}{\sum_{j \in E(\alpha)} w(\alpha, j)}
\end{eqnarray*}


\section{Experiments}
\label{sec:experiments} We conducted experiments to compare the
performance of Simrank, evidence-based Simrank and weighted
Simrank as techniques for query rewriting. Our baseline was a
query rewriting technique based on the Pearson correlation.
\subsection{Baseline}
The Pearson correlation between two queries $q$ and $q'$ is
defined as:
\begin{equation}
\textrm{sim}_{\textrm{pearson}} (q, q') = \frac{\sum_{\alpha \in
E(q)\bigcap E(q')} (w(q, \alpha) - \overline{w}_{q}) (w(q',
\alpha) -\overline{w}_{q'})}{\sqrt{\sum_{\alpha \in E(q)\bigcap
E(q')} (w(q, \alpha) - \overline{w}_{q})^2w(q', \alpha)
-\overline{w}_{q'})^2}}\nonumber
\end{equation}
where $\overline{w}_{q} = \sum_{i \in E(q)} \frac{w(q,
i)}{|E(q)|}$ is the average weight of all edges that have q as an
endpoint. If $E(q) \bigcap E(q') = \O$ then
$\textrm{sim}_{\textrm{pearson}}(q,q') = 0$. The Pearson
correlation indicates the strength of a linear relationship
between two variables. In our case, we use it to measure the
relationship between two queries. Notice, that
$\textrm{sim}_{\textrm{pearson}}$ takes values in the interval
$[-1, 1]$ and it requires that the two queries $q$ and $q'$ have
at least one common neighbor in the click graph.

\subsection{Dataset}
\label{sec:dataset} We started from a two-weeks click graph from
US Yahoo! search, containing approximately 15
million 
distinct queries, 14 million 
distinct ads and 28 million 
edges. An edge in this graph connects a query with an ad if and
only if the ad had been clicked at least once from a user that
issued the query. In addition, each edge contains the number of
clicks, the number of impressions, as well as the expected click
rate. This graph consists of one huge connected component and
several smaller subgraphs. In all our experiments that required
the use of an edge weight we used the expected click rate.

 To make the dataset size more manageable, we used the
subgraph extraction method described in \cite{DBLP:confs/ACL06} to
further decompose the largest component and we produced  five
smaller subgraphs. In summary, the algorithm in
\cite{DBLP:confs/ACL06} is an efficient local graph partitioning
algorithm that uses the PageRank vectors. Given a graph and an
initial node, it tries to find a cut with small conductance
\footnote{The conductance is a way to measure how hard it is to
leave a small set of a graph's nodes. If $\Phi_S$ is the
conditional probability of leaving a set of nodes $S$ given that
we started from a node in $S$, then the conductance is defined as
the minimal $\Phi_S$ over all sets $S$ that have a total
stationary probability of at most 1/2. More information can be
found in \cite{S93}.} near that starting node. We started from
different nodes and run the algorithm iteratively in order to
discover big enough, distinct subgraphs.
Table~\ref{table:fivesubgraphsdataset} tabulates the total number
of nodes (queries and ads) and edges contained in the
five-subgraphs dataset. We also observed a number of power-law
distributions, including ads-per-query, queries-per-ad and number
of clicks per query-ad pair. We used this dataset as the input
click graph for all query rewriting techniques we experimented
with.
\begin{table}
\begin{center}
\caption{Dataset statistics}
\begin{tabular} {crrr}
\hline
  & \# of Queries & \# of Ads &  \# of Edges\\
\hline subgraph 1 & 585,218 & 434,938 & 1,280,920\\
subgraph 2 & 530,797 & 374,243 & 1,130,314\\
subgraph 3 & 322,252 & 214,952 & 713,253\\
subgraph 4 & 313,951 & 243,406 & 703,747\\
subgraph 5 & 91,195 & 87,442 & 216,828\\
\hline Total &  1,843,413 & 1,354,981 & 4,045,062\\
 \hline
\end{tabular}
\label{table:fivesubgraphsdataset}
\end{center}
\end{table}

The query set for evaluation is sampled, with uniform probability,
from live traffic during the same two-weeks period. This traffic
contains all queries issued at Yahoo! during that period; even the
ones that did not bring any clicks on a sponsored search result.
More specifically, we used a standardized 1200 query sample that
has been generated by the above procedure and is currently being
used as a benchmark at Yahoo!. We looked at these 1200 queries and
extracted only the ones that actually appear in our five-subgraphs
dataset as only for those our query rewriting methods would be
able to provide rewrites. We found out that these are 120 queries
and these are the queries that constitute our evaluation set.
Using such an evaluation query selection procedure we made sure
that queries issued rarely had a smaller probability of appearing
in the evaluation set whereas more popular queries could appear
with higher probability. We made this decision since we are
interested in comparing the query rewriting techniques using a
realistic query set. In other words, we prefer a rewriting
technique that provides high quality rewrites for popular queries
from another one that does the same only for rare queries.

\subsection{Evaluation Method}
\label{sec:evaluation_method}
 We run each method on the five-subgraphs
dataset and recorded the top 100 rewrites for each query on our
queries sample. We then use stemming to filter out duplicate
rewrites (notice that such rewrites might appear in the click
graph). In addition we perform {\em bid term filtering}, i.e., we
remove queries that are {\em not} in a list of all queries that
saw bids in the two-week period when the click graph was gathered.
This list contains any query that received at least one bid at any
point in the period; hence, if a query is not in the list it is
unlikely to have bids currently. (Note that such queries with no
bids may still be connected to ads in the click graph. These ads
were displayed and clicked on because of query rewriting that took
place when the query was originally submitted.)

The queries that remain after duplicate elimination and bid term
filtering are considered for our evaluation. However, we limit
ourselves to at most 5 rewrites per query per method because of
the cost of the manual evaluation we describe next. Note that a
method may generate fewer than 5 rewrites after filtering. We call
the number of remaining rewrites the {\em depth} of a method.

To evaluate the quality of rewrites, we consider two methods. The
first is a manual evaluation, carried out by professional members
of Yahoo!'s editorial evaluation team. Each query -- rewrite pair
is considered by an evaluator, and is given a score on a scale
from 1 to 4, based on their relevance judgment. (The scoring is
the same as used in~\cite{DBLP:confs/JRMG2007,DBLP:confs/ZJ07}).
The query rewrites that were more relevant with the original query
assigned a score of 1, and the least related assigned a score of
4. Table \ref{table:editorial_scoring_system} summarizes the
interpretation of the four grades and their description is shown
below.

\begin{itemize}
\item[1.] Precise rewrite: The query rewrite matches the user's
intent and it preserves the core meaning of the original query
\item[2.] Approximate rewrite: The query rewrite has a direct
close relationship to the topic described by the initial query,
but the scope has narrowed or broadened or there has been a slight
shift to a closely related topic.
 \item[3.] Possible rewrite: The
query rewrite either has some categorical relationship to the
initial query (i.e. the two are in the same broad category of
products or services) or describes a complementary product, but is
otherwise distinct from the original user intent.
\item[4.] Clear
Mismatch: The query rewrite has no clear relationship to the
intent of the original query.
\end{itemize}

The judgment scores are solely based on the evaluator's knowledge,
and not on the contents of the click graph. Our second evaluation
method addresses the question of whether our methods made the
``right'' decision based on the evidence found in the click graph.
The basic idea is to remove certain edges from the click graph and
to see if using the remaining data our schemes can still make
useful inferences related to the missing data.

\begin{figure}[htbp]
\begin{center}
\includegraphics[scale=.55,angle=0]{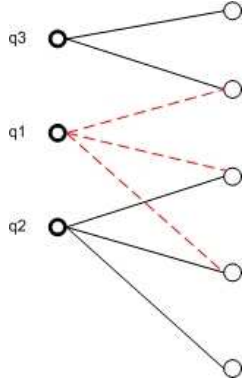}
 \caption{Sample setup for testing the ability of a rewriting method to compute correct query rewrites. By removing the red, dashed edges, we remove all direct similarity evidence
 between $q_1$ and $q_2, q_3$.}
\label{fig:desirability_example}
\end{center}
\end{figure}

In particular, consider Figure~\ref{fig:desirability_example},
showing two queries $q_2$ and $q_3$ that share at least one common
arc with a query $q_1$. In order to distinguish which query
between $q_2$ and $q_3$ is a preferable rewrite for $q_1$, we
define the desirability of query $q_2$ for query $q_1$ as
$\textrm{des}(q_1, q_2) = \sum_{i \in E(q_1)\bigcap E(q_2)}
\frac{1}{|E(q_2)|}\cdot w(q_2, i)$. By computing the desirability
scores $\textrm{des}(q_1, q_2), \textrm{des}(q_1, q_3)$ we can
determine the most desirable rewrite for $q_1$. That is, given the
evidence in the graph, if $\textrm{des}(q_1, q_2) >
\textrm{des}(q_1, q_3)$ then $q_2$ would be a better rewrite for
$q_1$ than $q_3$.

Given our definition of desirability, we can now conduct the
following experiment. First, we remove the edges that connect
$q_1$ to ads that are also connected with $q_3$ or $q_2$. In
Figure~\ref{fig:desirability_example} these are the red, dashed
edges. Then, we run each variation of Simrank on the remaining
graph and record the similarity scores $\textrm{sim}(q_1, q_2)$
and $\textrm{sim}(q_1, q_3)$ that the method gives. Finally, we
test whether the ordering for $q_2, q_3$ that these similarity
scores provide is consistent with the ordering derived from the
desirability scores. In our example, if $\textrm{des}(q_1, q_2)
> \textrm{des}(q_1, q_3)$ and $\textrm{sim}(q_1, q_2) > \textrm{sim}(q_1,
q_3)$ then we would say that the similarity score was successful
in predicting the desirable rewrite.

We repeated this edge removal experiment for $50$ queries randomly
selected from our five-subgraphs dataset. These queries played the
role of query $q_1$ as described above. For each of those queries
we identified all the queries from the dataset that shared at
least one common ad with it and we randomly selected two of them.
Those were the $q_2$ and $q_3$ queries. In order to make sure that
a Simrank similarity score can be computed after the deletion of
the edges in our experiment, we selected the queries $q_2, q_3$
after making sure that after edge removal there would still exist
a path from $q_2$ to $q_1$ and from $q_3$ to $q_1$ through other
edges in the graph. Since Pearson correlation only can be used
provided that there is at least a common ad between two queries,
we did not include the technique in this part of our evaluation.

\subsection{Metrics}
\begin{table}
\begin{small}
\begin{center}
 \caption{Editorial scoring system for query rewrites.}
\begin{tabular} {|l|l|l|l|}
\hline
Score &  & Definition &  Example (query - re-write)\\
\hline 1 & Precise Match & near-certain match & corvette car -
chevrolet corvette \\
\hline 2 & Approximate Match & probable, but inexact match with user intent & apple music player - ipod shuffle \\
\hline 3 & Marginal Match & distant, but plausible match to a
related topic & glasses - contact lenses \\
\hline 4 & Mismatch & clear mismatch & time magazine - time \&
date magazine\\
 \hline
\end{tabular}
\label{table:editorial_scoring_system}
\end{center}
\end{small}
\end{table}
The evaluation metrics we used were the following four:
\begin{itemize}
\item[(i)] Precision/recall: We consider two IR tasks. Firstly, we
interpret the rewrites with scores 1-2 as relevant queries and the
rewrites with scores 3-4 as irrelevant queries. Secondly, we
interpret as relevant query rewrites only the ones with score 1
and the rest as irrelevant. Thus, we can define the
precision/recall of method $m$ for query $q$ as follows:
\begin{eqnarray*}
\textrm{precision}(q, m)& =& \frac{\textrm{relevant rewrites of
$q$ that $m$ provides}}{\textrm{number of rewrites for $q$ that
$m$ provides}}\\
 \textrm{recall}(q, m)& =& \frac{\textrm{relevant
rewrites of $q$ that $m$ provides}}{\textrm{number of relevant
rewrites for $q$ among all methods}}
\end{eqnarray*}
\item[(ii)] Query Coverage: We are also interested in the absolute
number of queries (from our 120 query sample) for which each
method manages to provide at least one rewrite. We call this
number query coverage. In general, we prefer methods that cover as
many as possible queries.

\item[(iii)] Query rewriting depth: Here, we are interested in the
total number of query rewrites that a method provides for a given
query. This is called the depth of a query rewriting technique.
Again, we are interested in methods that have larger rewriting
depth.

\item[(iv)] Desirability prediction: For our desirability
experiment, we report the fraction of the $50$ queries for which a
method was able to correctly predict the desirability of $q_2$ (or
$q_3$) over the other query.
\end{itemize}

\section{Results} \label{sec:results}
\subsection{Query Coverage}
\begin{figure}[htbp]
 \begin{center}
\includegraphics[scale=.60,angle=0]{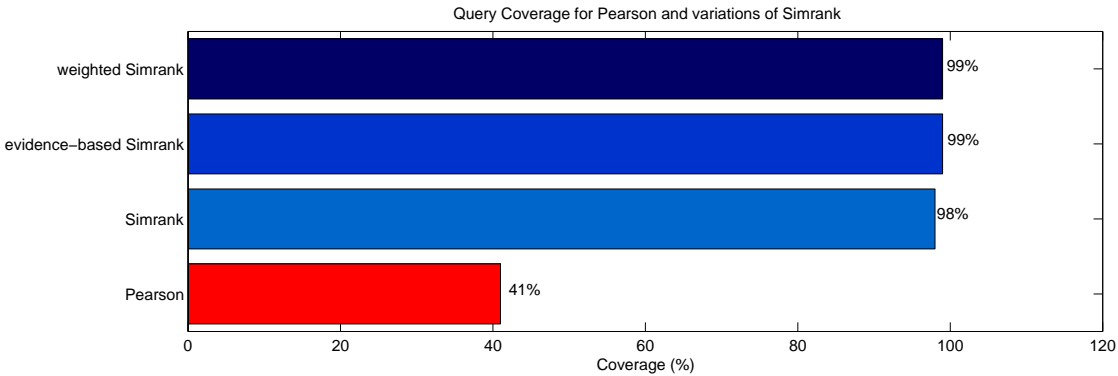}
 \caption{Comparing the query coverage of Pearson and Simrank}
\label{fig:coverage_comparison_simrank_pearson}
\end{center}
\end{figure}
Figure \ref{fig:coverage_comparison_simrank_pearson} illustrates
the percentage of queries from the 120  queries sample that
Pearson and Simrank provide rewrites for. Simrank provides
rewrites almost for all queries (98\%) when Pearson gives rewrites
only for the 41\% of the queries. This can be considered as
expected, since Pearson can only measure similarity between two
queries if they share a common ad, whereas Simrank takes into
account the whole graph structure and does not require something
similar. Also notice, that evidence-based Simrank further improves
the coverage to 99\%.

\subsection{Precision-Recall}
\begin{figure}[htbp]
\begin{center}
\includegraphics[scale=.55,angle=0]{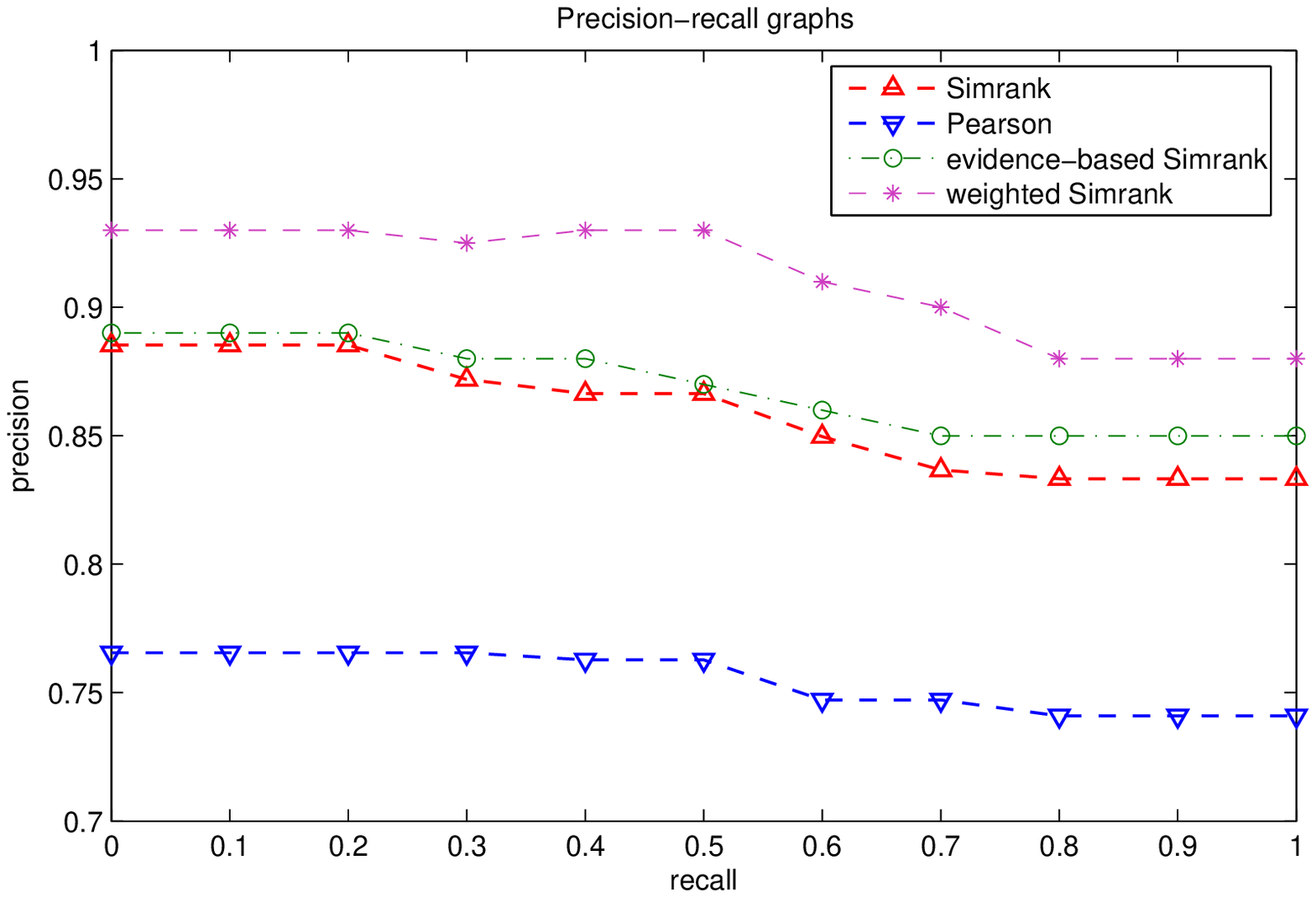}
\includegraphics[scale=.55,angle=0]{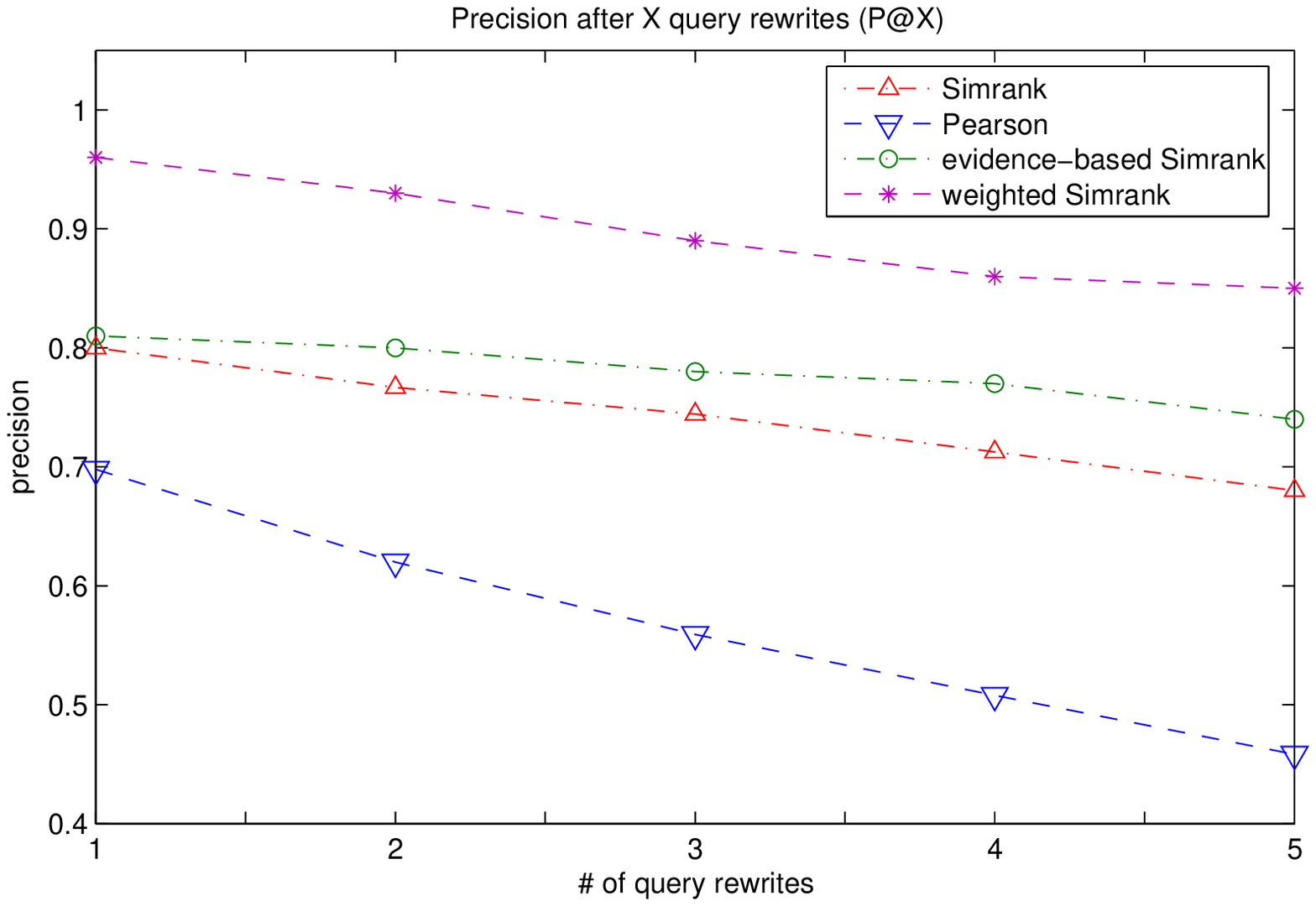}
 \caption{Precision at 11 standard
recall levels (top) and precision after $X = 1,2,\dots, 5$ query
rewrites (P@X) (bottom) using as positive class rewrites with
score \{1-2\} and negative class rewrites with score \{3-4\}}
\label{fig:precision_recall_threshold_2}
\end{center}
\end{figure}

\begin{figure}[]
\begin{center}
\includegraphics[scale=.55,angle=0]{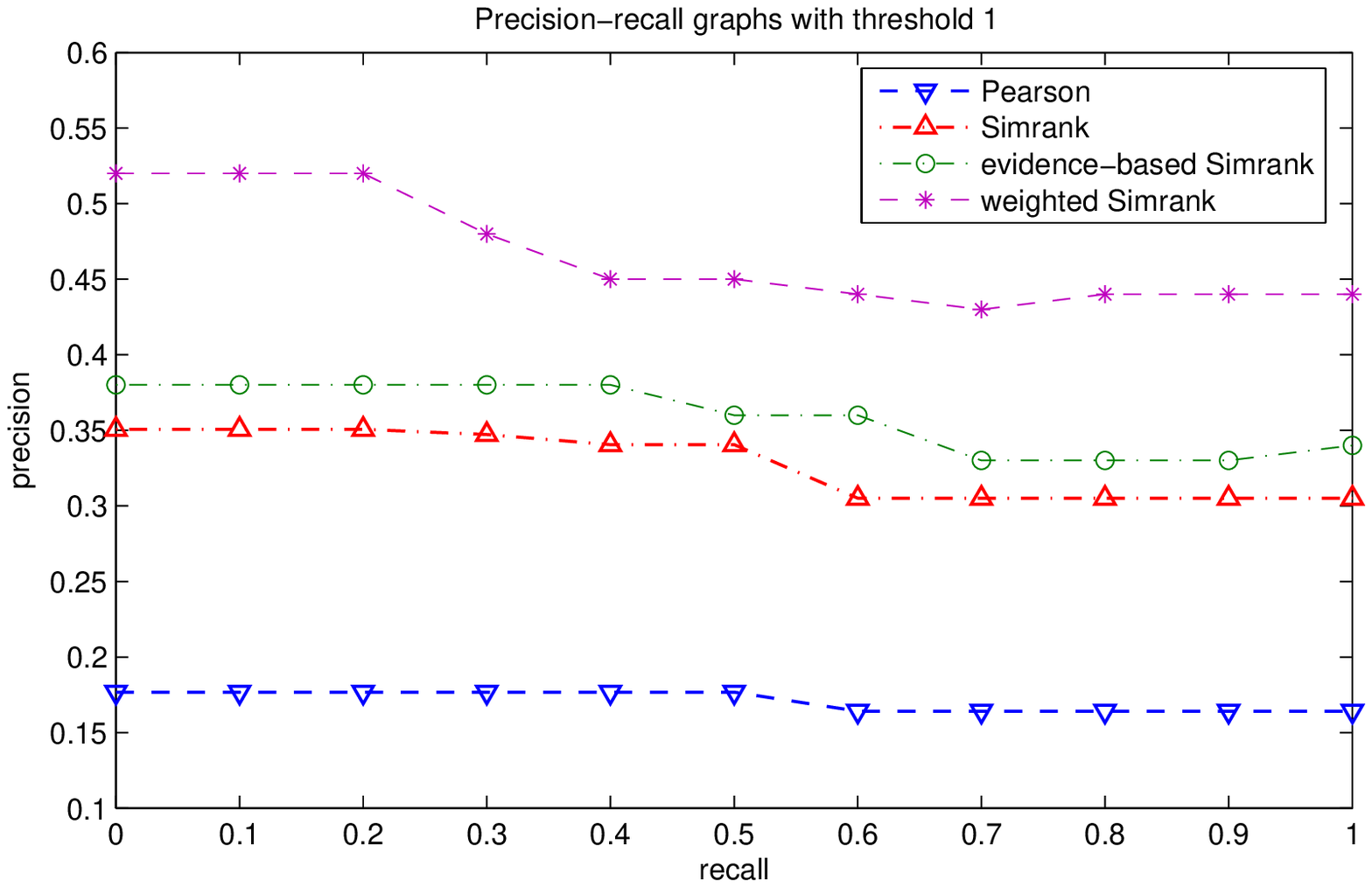}
\includegraphics[scale=.55,angle=0]{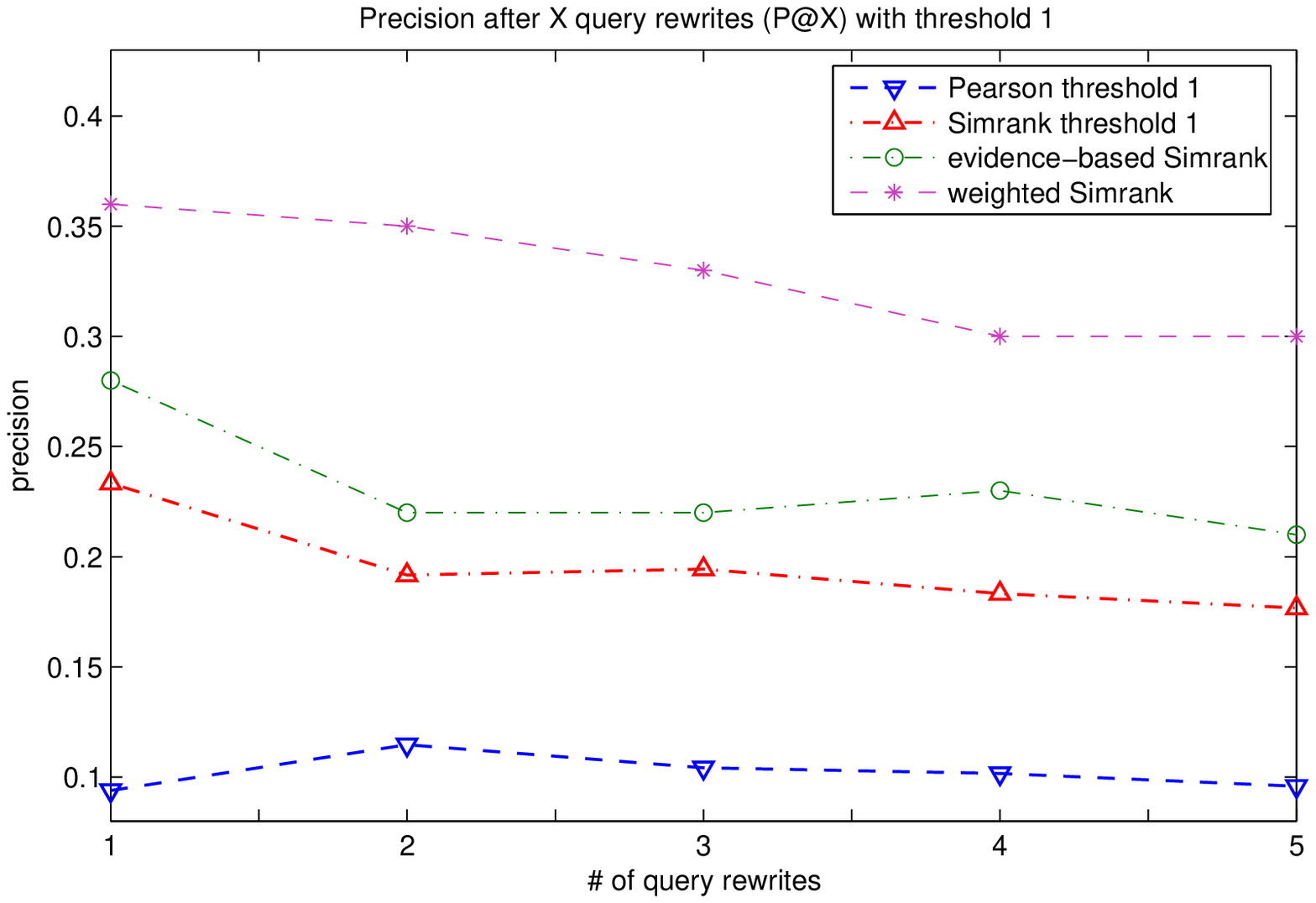}
 \caption{Precision at 11 standard
recall levels (top) and precision after $X = 1,2,\dots, 5$ query
rewrites (P@X) (bottom) using as positive class rewrites with
score 1 and negative class rewrites with score
\{2-4\}}\label{fig:precision_recall_threshold_1}
\end{center}
\end{figure}

Figure~\ref{fig:precision_recall_threshold_2} presents the
precision/recall graphs for Pearson and Simrank as well as the
precision at 1-5 queries (P@X). For the computation of precision
and recall the editorial scores were used in a binary
classification manner; scores 1-2 were the positive class and
scores 3-4 the negative class. For instance, in
Figure~\ref{fig:precision_recall_threshold_2} (bottom graph) we
see that Weighted Simrank has 93\% precision for 2 rewrites,
meaning that 93\% of its rewrites in the top two ranks were given
scores of 1 or 2 by the evaluators. Figure
\ref{fig:precision_recall_threshold_1} presents more
precision/recall graphs for Pearson and Simrank as well as
precision at 1-5 queries (P@X). However, now, the positive class
of the binary classification problem consists of the editorial
score 1, whereas the negative class contains the editorial scores
2-4.

In both cases we see that simple Simrank substantially improves
the precision of the rewrites compared to Pearson. In addition,
the use of the evidence score and the exploitation of the graph
weights further boosts the precision, as expected.

\subsection{Rewriting Depth}
\begin{figure}[htbp]
 \begin{center}
\includegraphics[scale=.70,angle=0]{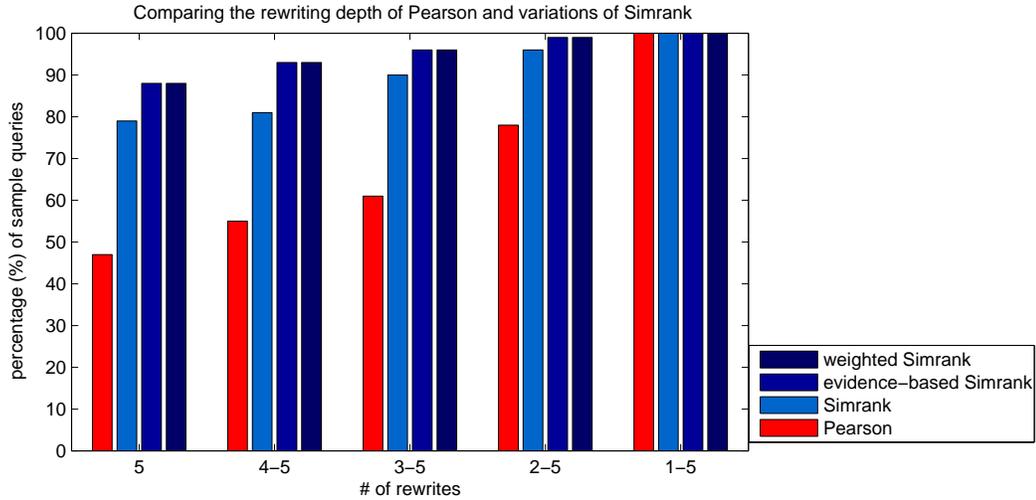}
 \caption{Comparing the rewriting depth of Pearson and Simrank}
\label{fig:depth_comparison_simrank_pearson}
\end{center}
\end{figure}

Figure \ref{fig:depth_comparison_simrank_pearson} compares the
rewriting depth of Pearson and the variations of Simrank. Note
that our two enhanced schemes can provide the full 5 rewrites for
over 85\% of the queries. As mentioned earlier, the more rewrites
we can generate, the more options the back-end will have for
finding ads with active bids.

\subsection{Desirability prediction}
\begin{figure}[htbp]
 \begin{center}
\includegraphics[scale=.60,angle=0]{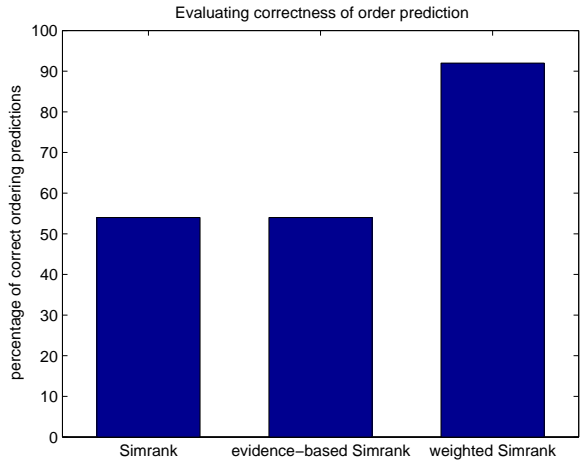}
 \caption{Comparing the ability of query rewriting methods to correctly predict the order of query rewrite candidates}
\label{fig:desirability}
\end{center}
\end{figure}
Figure \ref{fig:desirability} provides the results of our
experiments for identifying the correct order of query rewrites as
described in Section \ref{sec:evaluation_method}. Simple Simrank
and evidence-based Simrank manage to predict successfully the
desirable rewrite for 27 out of the 50 queries (54\%). Note that
both methods do not exploit the graph weights in the similarity
computations and rely only on the graph structure. Weighted
Simrank predicts correctly the desirable rewrite for 46 queries
(92\%).

\subsection{Discussion}
As we can see, simple Simrank outperforms Pearson both in query
coverage, rewriting depth and precision/recall. Notice here that
this version of Simrank does not utilize at all the qualitative
information in the click graph, whereas Pearson does.

The introduction of evidence scores increases query coverage
slightly (by 1\%) and substantially improves the quality of the
rewrites. For instance, the precision at 5 rewrites of simple
Simrank is 75\% whereas the precision after 5 rewrites of
evidence-based Simrank is 80\% (Figure
\ref{fig:precision_recall_threshold_2}). In addition, in the P@X
rewrites diagram (Figure \ref{fig:precision_recall_threshold_2})
the line corresponding to the precision of evidence-based Simrank
is always above the one corresponding to the precision of simple
Simrank. Finally,  evidence-based Simrank increases the rewriting
depth. For example, simple Simrank provides five rewrites for 79\%
of the queries, whereas evidence based Simrank gives five rewrites
for the 89\% of the queries (Figure
\ref{fig:depth_comparison_simrank_pearson}).

Weighted Simrank builds upon evidence-based Simrank and utilizes
the graph weights. It maintains the query coverage percentage of
evidence-based Simrank at 99\% (Figure
\ref{fig:coverage_comparison_simrank_pearson}) and  substantially
improves the quality of the rewrites.  Figure
\ref{fig:precision_recall_threshold_2} shows that the P@X line of
weighted Simrank is always above the one of evidence-based
Simrank. The precision at 5 rewrites of weighted Simrank goes from
80\% (evidence-based Simrank) to 86\%. Also, 96\% of the queries
have a high-quality top rewrite when we use weighted Simrank (P@1,
Figure \ref{fig:precision_recall_threshold_2}) when the
corresponding percentages for evidence-based Simrank, simple
Simrank and Pearson are 81\%, 80\% and 70\%. In our desirability
experiment, weighted Simrank predicted successfully the desirable
rewrite for 92\% of the cases (Figure \ref{fig:desirability}).
Finally, weighted-based Simrank maintains the rewriting depth of
evidence-based Simrank (Figure
\ref{fig:depth_comparison_simrank_pearson}).

\section{Conclusions}
\label{sec:conclusions}

In this paper we focused on the problem of query rewriting for
sponsored search. We proposed Simrank to exploit the click graph
structure and we introduced two extensions: one that takes into
account the weights of the edges in the click graph, and another
that takes into account the ``evidence'' supporting the similarity
between queries. Our experimental results show that weighted-based
Simrank is the overall best method for generating rewrites based
on a click graph.

There are several query rewriting issues that we did not address
in our analysis. Spam clicks can mislead our techniques and thus
spam-resistant variations of our techniques would be useful. Also,
methods for combining our similarity scores with semantic
text-based similarities could be considered.

Even though our new schemes were developed and tested for query
rewriting based on a click graph, we suspect that the weighted and
evidence-based Simrank methods could be of use in other
applications that exploit bi-partite graphs. We plan to experiment
with these schemes in other domains, including collaborative
filtering.

\section{Acknowledgements}

We thank Kevin Lang for providing us his code from
\cite{DBLP:confs/ACL06} and for helping us with the subgraph
extraction procedure. We also thank Yahoo! Editorial team for
carrying out the evaluation of our techniques. Finally, we thank
Panagiotis Papadimitriou, Zolt\'{a}n Gy{$\ddot{\mbox{o}}$}ngyi,
Tasos Anastasakos and Tam\'{a}s Sarl\'{o}s for fruitful
discussions.
\bibliography{references}

\appendix

\newpage
\section{Simrank similarity scores on complete bipartite graphs}
\label{sec:appendix1}

\begin{theorem}
Consider the complete bipartite graph $K_{2,2}$ with nodes sets
$V_1 = \{a, b\}$ and $V_2 = \{ A, B\}$. Let $\textrm{sim}^{(k)}
(A,B)$ denote the similarity between nodes $A, B$ that bipartite
Simrank computes after $k$ iterations and let $C_1, C_2$ denote
the Simrank decay factors. Then:

\begin{itemize}
\item[(i)]$\textrm{sim}^{(k)} (A,B) = \frac{C_2}{2}\sum_{i=1}^{k}
\frac{1}{2^{i-1}}C_1^{\lfloor\frac{i}{2}\rfloor}
C_2^{\lceil\frac{i-1}{2}\rceil}$

\item[(ii)] $\lim_{k \rightarrow \infty}\textrm{sim}^{(k)} (A,B)
\leq C_2$
\end{itemize}

\begin{proof}
(i) We will follow the computation of the Simrank similarity
scores from equations \ref{eq:query_similarities} and
\ref{eq:ad_similarities}.
\begin{itemize}

\item Iteration 1:
\begin{eqnarray}
\textrm{sim}^{(1)}(A, B)& =& \frac{C_2}{2\cdot 2} \left(1+1\right)\nonumber\\
& =& \frac{C_2}{2}\nonumber\\
& = & \frac{C_2}{2}\sum_{i=1}^{1}
\frac{1}{2^{i-1}}C_1^{\lfloor\frac{i}{2}\rfloor}
C_2^{\lceil\frac{i-1}{2}\rceil} \nonumber
\end{eqnarray}
\begin{eqnarray}
\textrm{sim}^{(1)}(a, b) &=
&\frac{C_1}{2\cdot 2} \left(1+1\right) \nonumber\\
 &=& \frac{C_1}{2}\nonumber
\end{eqnarray}

\item Iteration 2:
\begin{eqnarray}
\textrm{sim}^{(2)}(A, B) &=& \frac{C_2}{2\cdot 2} \left(1+1 +
\frac{C_1}{2}+
\frac{C_1}{2}\right)\nonumber\\
& =& \frac{C_2}{2} + \frac{C_1 \cdot C_2}{4}\nonumber\\
& =& \frac{C_2}{2} \left(1+\frac{C_1}{2}\right)\nonumber\\
& = & \frac{C_2}{2}\sum_{i=1}^{2}
\frac{1}{2^{i-1}}C_1^{\lfloor\frac{i}{2}\rfloor}
C_2^{\lceil\frac{i-1}{2}\rceil} \nonumber
\end{eqnarray}
\begin{eqnarray}
\textrm{sim}^{(2)}(a, b) &= &\frac{C_1}{2\cdot 2} \left(1+1 +
\frac{C_2}{2}+
\frac{C_2}{2}\right) \nonumber \\
&= &\frac{C_1}{2} + \frac{C_1 \cdot C_2}{4}\nonumber \\
& = &\frac{C_1}{2} \left(1+\frac{C_2}{2}\right)\nonumber
\end{eqnarray}

\item Iteration 3:
\begin{eqnarray}
\textrm{sim}^{(3)}(A, B)& =& \frac{C_2}{2\cdot 2} \left(1+1 +
\left(\frac{C_1}{2}+ \frac{C_1 \cdot C_2}{4}\right)  + \left(\frac{C_1}{2}+ \frac{C_1 \cdot C_2}{4}\right) \right)\nonumber \\
&= &\frac{C_2}{2} + \frac{C_1 \cdot C_2}{4} + \frac{C_2^2 \cdot
C_1}{8} \nonumber \\
&= &\frac{C_2}{2} \left(1+\frac{C_1}{2} + \frac{C_2 \cdot
C_1}{4}\right)\nonumber\\
& = & \frac{C_2}{2}\sum_{i=1}^{3}
\frac{1}{2^{i-1}}C_1^{\lfloor\frac{i}{2}\rfloor}
C_2^{\lceil\frac{i-1}{2}\rceil} \nonumber
\end{eqnarray}
\begin{eqnarray}
\textrm{sim}^{(3)}(a, b) &=& \frac{C_1}{2\cdot 2} \left(1+1 +
\left(\frac{C_2}{2}+ \frac{C_1 \cdot C_2}{4}\right) + \left(\frac{C_2}{2}+ \frac{C_1 \cdot C_2}{4}\right) \right)\nonumber\\
& =& \frac{C_1}{2} + \frac{C_1 \cdot C_2}{4} + \frac{C_1^2 \cdot
C_2}{8} \nonumber \\
&= &\frac{C_1}{2} \left(1+\frac{C_2}{2} + \frac{C_1 \cdot
C_2}{4}\right)\nonumber
\end{eqnarray}

\item Iteration 4:
\begin{eqnarray}
\textrm{sim}^{(4)}(A, B)& =& \frac{C_2}{2\cdot 2} \left(1+1 +
\left(\frac{C_1}{2}+ \frac{C_1 \cdot C_2}{4} + \frac{C_1^2 \cdot C_2}{8}\right) + \left(\frac{C_1}{2}+ \frac{C_1 \cdot C_2}{4} + \frac{C_1^2 \cdot C_2}{8}\right) \right) \nonumber \\
&= &\frac{C_2}{2} + \frac{C_1 \cdot C_2}{4} + \frac{C_2^2 \cdot
C_1}{8} + \frac{C_2^2 \cdot C_1^2}{16} \nonumber \\
&= &\frac{C_2}{2} \left(1+\frac{C_1}{2} + \frac{C_2 \cdot C_1}{4}
+ \frac{C_1^2 \cdot C_2}{8}\right)\nonumber\\
& = & \frac{C_2}{2}\sum_{i=1}^{4}
\frac{1}{2^{i-1}}C_1^{\lfloor\frac{i}{2}\rfloor}
C_2^{\lceil\frac{i-1}{2}\rceil} \nonumber
\end{eqnarray}
\begin{eqnarray}
\textrm{sim}^{(4)}(a, b)& =& \frac{C_1}{2\cdot 2} \left(1+1 +
\left(\frac{C_2}{2}+ \frac{C_1 \cdot C_2}{4} + \frac{C_2^2 \cdot C_1}{8}\right) + \left(\frac{C_2}{2}+ \frac{C_1 \cdot C_2}{4} + \frac{C_2^2 \cdot C_1}{8}\right) \right) \nonumber \\
&= &\frac{C_1}{2} + \frac{C_1 \cdot C_2}{4} + \frac{C_1^2 \cdot
C_2}{8} + \frac{C_2^2 \cdot C_1^2}{16} \nonumber \\
&= &\frac{C_1}{2} \left(1+\frac{C_2}{2} + \frac{C_2 \cdot C_1}{4}
+ \frac{C_2^2 \cdot C_1}{8}\right)\nonumber
\end{eqnarray}

\item Iteration 5:
\begin{eqnarray}
\textrm{sim}^{(5)}(A, B)& =& \frac{C_2}{2\cdot 2} \left(1+1 +
\left(\frac{C_1}{2}+ \frac{C_1 \cdot C_2}{4} + \frac{C_1^2 \cdot C_2}{8} + \frac{C_1^2 \cdot C_2^2}{16}\right) + \left(\frac{C_1}{2}+ \frac{C_1 \cdot C_2}{4} + \frac{C_1^2 \cdot C_2}{8}+ \frac{C_1^2 \cdot C_2^2}{16}\right) \right) \nonumber \\
&= &\frac{C_2}{2} + \frac{C_1 \cdot C_2}{4} + \frac{C_2^2 \cdot
C_1}{8} + \frac{C_2^2 \cdot C_1^2}{16} + \frac{C_1^2 \cdot C_2^3}{32} \nonumber \\
&= &\frac{C_2}{2} \left(1+\frac{C_1}{2} + \frac{C_2 \cdot C_1}{4}
+ \frac{C_1^2 \cdot C_2}{8} + \frac{C_1^2 \cdot
C_2^2}{16}\right)\nonumber\\
& = & \frac{C_2}{2}\sum_{i=1}^{5}
\frac{1}{2^{i-1}}C_1^{\lfloor\frac{i}{2}\rfloor}
C_2^{\lceil\frac{i-1}{2}\rceil} \nonumber
\end{eqnarray}
\begin{eqnarray}
\textrm{sim}^{(5)}(a, b)& =& \frac{C_1}{2\cdot 2} \left(1+1 +
\left(\frac{C_2}{2}+ \frac{C_1 \cdot C_2}{4} + \frac{C_2^2 \cdot C_1}{8} + \frac{C_1^2 \cdot C_2^2}{16}\right) + \left(\frac{C_2}{2}+ \frac{C_1 \cdot C_2}{4} + \frac{C_2^2 \cdot C_1}{8}+ \frac{C_1^2 \cdot C_2^2}{16}\right) \right) \nonumber \\
&= &\frac{C_1}{2} + \frac{C_1 \cdot C_2}{4} + \frac{C_1^2 \cdot
C_2}{8} + \frac{C_2^2 \cdot C_1^2}{16} + \frac{C_2^2 \cdot C_1^3}{32} \nonumber \\
&= &\frac{C_1}{2} \left(1+\frac{C_2}{2} + \frac{C_2 \cdot C_1}{4}
+ \frac{C_2^2 \cdot C_1}{8} + \frac{C_1^2 \cdot
C_2^2}{16}\right)\nonumber
\end{eqnarray}
\item Iteration 6:
\begin{eqnarray}
\textrm{sim}^{(6)}(A, B)& =& \frac{C_2}{2\cdot 2} \left(1+1 +
2 \cdot \left(\frac{C_1}{2}+ \frac{C_1 \cdot C_2}{4} + \frac{C_1^2 \cdot C_2}{8} + \frac{C_1^2 \cdot C_2^2}{16} + \frac{C_1^3 \cdot C_2^2}{32}\right)\right) \nonumber\\
&= &\frac{C_2}{2} + \frac{C_1 \cdot C_2}{4} + \frac{C_2^2 \cdot
C_1}{8} + \frac{C_2^2 \cdot C_1^2}{16} + \frac{C_1^2 \cdot C_2^3}{32} + \frac{C_1^3 \cdot C_2 ^ 3}{64} \nonumber \\
&= &\frac{C_2}{2} \left(1+\frac{C_1}{2} + \frac{C_2 \cdot C_1}{4}
+ \frac{C_1^2 \cdot C_2}{8} + \frac{C_1^2 \cdot C_2^2}{16} +
\frac{C_1^3 \cdot C_2^2}{32}\right)\nonumber\\
& = & \frac{C_2}{2}\sum_{i=1}^{6}
\frac{1}{2^{i-1}}C_1^{\lfloor\frac{i}{2}\rfloor}
C_2^{\lceil\frac{i-1}{2}\rceil} \nonumber
\end{eqnarray}
\begin{eqnarray}
\textrm{sim}^{(6)}(a, b)& =& \frac{C_1}{2\cdot 2} \left(1+1 +
2 \cdot \left(\frac{C_2}{2}+ \frac{C_1 \cdot C_2}{4} + \frac{C_2^2 \cdot C_1}{8} + \frac{C_1^2 \cdot C_2^2}{16} + \frac{C_2^3 \cdot C_1^2}{32}\right) \right) \nonumber \\
&= &\frac{C_1}{2} + \frac{C_1 \cdot C_2}{4} + \frac{C_1^2 \cdot
C_2}{8} + \frac{C_2^2 \cdot C_1^2}{16} + \frac{C_2^2 \cdot C_1^3}{32} + \frac{C_1^3 \cdot C_2 ^ 3}{64} \nonumber \\
&= &\frac{C_1}{2} \left(1+\frac{C_2}{2} + \frac{C_2 \cdot C_1}{4}
+ \frac{C_2^2 \cdot C_1}{8} + \frac{C_1^2 \cdot C_2^2}{16} +
\frac{C_2^3 \cdot C_1^2}{32}\right)\nonumber
\end{eqnarray}
\end{itemize}

We can easily observe that $\textrm{sim}^{(k)} (A,B) =
\frac{C_2}{2}\sum_{i=1}^{k}
\frac{1}{2^{i-1}}C_1^{\lfloor\frac{i}{2}\rfloor}
C_2^{\lceil\frac{i-1}{2}\rceil}$ \\

(ii) We know that $C_1  \leq 1$ and $C_2 \leq 1$. Thus:
\begin{equation*}
\frac{C_2}{2}\sum_{i=1}^{k}
\frac{1}{i}C_1^{\lfloor\frac{i}{2}\rfloor}
C_2^{\lceil\frac{i-1}{2}\rceil} \leq \frac{C_2}{2}\sum_{i=1}^{k}
\frac{1}{2^{i-1}}
\end{equation*}

Now we can write:

\begin{eqnarray*}
\lim_{k \rightarrow \infty}\textrm{sim}^{(k)} (A,B)  &= &\lim_{k
\rightarrow \infty} \frac{C_2}{2}\sum_{i=1}^{k}
\frac{1}{2^{i-1}}C_1^{\lfloor\frac{i}{2}\rfloor}
C_2^{\lceil\frac{i-1}{2}\rceil}\\
& \leq & \lim_{k \rightarrow \infty}\frac{C_2}{2}\sum_{i=1}^{k}
\frac{1}{2^{i-1}}  = \frac{C_2}{2} \lim_{k \rightarrow \infty}
\sum_{i=1}^{k} \frac{1}{i}=  \frac{C_2}{2} \cdot 2  =  C_2
\end{eqnarray*}

Thus,  $\lim_{k \rightarrow \infty}\textrm{sim}^{(k)} (A,B) \leq
C_2$. \qed
\end{proof}

\label{theorem1}
\end{theorem}

\begin{theorem}
Consider the two complete bipartite graphs $G = K_{1,2}$ and $G' =
K_{2,2}$ with nodes sets $V_1 = \{a\}, V_2 = \{A, B\}$ and $V'_1 =
\{b, c\}$ and $V'_2 = \{C,D\}$ correspondingly. Let
$\textrm{sim}^{(k)}(A, B)$ and $\textrm{sim}^{(k)}(C, D)$ denote
the similarity scores that bipartite Simrank  computes for the
node pairs $(A,B)$ and $(C, D)$ after $k$ iterations. Then,
$\textrm{sim}^{(k)}(A, B) \geq \textrm{sim}^{(k)}(C,D),\ \forall \
k>0$.

\begin{proof}
From equations \ref{eq:query_similarities},
\ref{eq:ad_similarities} we have:
\begin{equation*}
\textrm{sim}^{(k)}(A, B) = \frac{C_2}{1 \cdot 1} 1 = C_2,\ \forall
k
> 0
\end{equation*}

Also, from Theorem \ref{theorem1}(i), we have:
\begin{equation*}
\textrm{sim}^{(k)}(C, D) = \frac{C_2}{2}\sum_{i=1}^{k}
\frac{1}{2^{i-1}}C_1^{\lfloor\frac{i}{2}\rfloor}
C_2^{\lceil\frac{i-1}{2}\rceil} \leq \lim_{k \rightarrow
\infty}\frac{C_2}{2}\sum_{i=1}^{k} \frac{1}{2^{i-1}} =
\frac{C_2}{2} \cdot 2 = C_2
\end{equation*}

Thus, $\textrm{sim}^{(k)}(A, B) \geq \textrm{sim}^{(k)}(C,D),\
\forall \ k>0$. \qed
\end{proof}
\label{theorem2}
\end{theorem}

\begin{theorem}
Consider the two complete bipartite graphs $G = K_{1,2}$ and $G' =
K_{2,2}$ with nodes sets $V_1 = \{a\}, V_2 = \{A, B\}$ and $V'_1 =
\{b, c\}$ and $V'_2 = \{C,D\}$ correspondingly. Let
$\textrm{sim}^{(k)}(A, B)$ and $\textrm{sim}^{(k)}(C, D)$ denote
the similarity scores that bipartite Simrank computes for the node
pais $(A,B)$ and $(C, D)$ after $k$ iterations. Then, $\lim_{k
\rightarrow \infty} \ \textrm{sim}^{(k)}(A, B) = \lim_{k
\rightarrow \infty}\ \textrm{sim}^{(k)}(C,D)$ if and only if $C_1
= C_2 = 1$, where $C_1,\ C_2$ are the decay factors of the
bipartite Simrank equations.

\begin{proof}
Let us assume that $\lim_{k \rightarrow \infty} \
\textrm{sim}^{(k)}(A, B) = \lim_{k \rightarrow \infty}\
\textrm{sim}^{(k)}(C,D)$.\\

That means that:
\begin{eqnarray*}
 \lim_{k \rightarrow \infty} \ \textrm{sim}^{(k)}(A,
B) &=&\lim_{k \rightarrow \infty}\ \textrm{sim}^{(k)}(C,D)  \Leftrightarrow  \\
C_2 & = & \lim_{k \rightarrow \infty} \frac{C_2}{2}\sum_{i=1}^{k}
\frac{1}{2^{i-1}}C_1^{\lfloor\frac{i}{2}\rfloor}
C_2^{\lceil\frac{i-1}{2}\rceil} \Leftrightarrow \\
C_2 & = & \frac{C_2}{2} \lim_{k \rightarrow \infty}
\frac{1}{2^{i-1}}C_1^{\lfloor\frac{i}{2}\rfloor}
C_2^{\lceil\frac{i-1}{2}\rceil} \Leftrightarrow \\
1 & = & \frac{1}{2} \lim_{k \rightarrow \infty}
\frac{1}{2^{i-1}}C_1^{\lfloor\frac{i}{2}\rfloor}
C_2^{\lceil\frac{i-1}{2}\rceil} \Leftrightarrow \\
\lim_{k \rightarrow \infty}
\frac{1}{2^{i-1}}C_1^{\lfloor\frac{i}{2}\rfloor}
C_2^{\lceil\frac{i-1}{2}\rceil} &= &2  \Leftrightarrow \\
C_1 & = &  C_2 = 1
\end{eqnarray*}

Now, let us assume that $C_1 = C_2 = 1$. We will have:

\begin{eqnarray*}
\lim_{k \rightarrow \infty}\ \textrm{sim}^{(k)}(C,D) & = & \lim_{k
\rightarrow \infty} \frac{C_2}{2}\sum_{i=1}^{k}
\frac{1}{2^{i-1}}C_1^{\lfloor\frac{i}{2}\rfloor}
C_2^{\lceil\frac{i-1}{2}\rceil}\\
& = &  \lim_{k \rightarrow \infty} \frac{C_2}{2}\sum_{i=1}^{k}
\frac{1}{2^{i-1}}\\
& = & C_2\\
& = & \lim_{k \rightarrow \infty}\ \textrm{sim}^{(k)}(A,B)
\end{eqnarray*}

Thus, $\lim_{k \rightarrow \infty} \ \textrm{sim}^{(k)}(A, B) =
\lim_{k \rightarrow \infty}\ \textrm{sim}^{(k)}(C,D)$ if and only
if $C_1 = C_2 = 1$. \qed
\end{proof}
\label{theorem3}
\end{theorem}

\begin{corollary}
Consider the two complete bipartite graphs $G = K_{1,2}$ and $G' =
K_{2,2}$ with nodes sets $V_1 = \{a\}, V_2 = \{A, B\}$ and $V'_1 =
\{b, c\}$ and $V'_2 = \{C,D\}$ correspondingly. Let
$\textrm{sim}^{(k)}(A, B)$ and $\textrm{sim}^{(k)}(C, D)$ denote
the similarity scores that bipartite Simrank computes for the node
pais $(A,B)$ and $(C, D)$ after $k$ iterations. Then, if for the
 decay factors of bipartite Simrank $C_1,\ C_2$ we know that $C_1 < 1$
 or $C_2 < 1$, both of the following are true:

\begin{itemize}

\item[(i)] $\textrm{sim}^{(k)}(A, B) > \textrm{sim}^{(k)}(C,D),\
\forall \ k>0$, and

\item[(ii)]   $\lim_{k \rightarrow \infty} \ \textrm{sim}^{(k)}(A,
B) > \lim_{k \rightarrow \infty}\ \textrm{sim}^{(k)}(C,D)$
\end{itemize}

\begin{proof}
\begin{itemize}
\item[(i)] It follows directly from Theorems \ref{theorem2} and
\ref{theorem3}. \item[(ii)] We have:
\begin{eqnarray*}
\lim_{k \rightarrow \infty}\ \textrm{sim}^{(k)}(C,D) & = & \lim_{k
\rightarrow \infty} \frac{C_2}{2}\sum_{i=1}^{k}
\frac{1}{2^{i-1}}C_1^{\lfloor\frac{i}{2}\rfloor}
C_2^{\lceil\frac{i-1}{2}\rceil}\\
& < &  \lim_{k \rightarrow \infty} \frac{C_2}{2}\sum_{i=1}^{k}
\frac{1}{2^{i-1}}\\
& = & C_2\\
& = & \lim_{k \rightarrow \infty}\ \textrm{sim}^{(k)}(A,B)
\end{eqnarray*}
Thus, $\lim_{k \rightarrow \infty} \ \textrm{sim}^{(k)}(A, B) >
\lim_{k \rightarrow \infty}\ \textrm{sim}^{(k)}(C,D)$. \qed
\end{itemize}
\end{proof}
\label{corollary1}
\end{corollary}

\begin{theorem}
Consider the two complete bipartite graphs $G = K_{m,2}$ and $G' =
K_{n,2}$ with $m < n$ and nodes sets $V_1, V_2 = \{A, B\}$ and
$V'_1, V'_2 = \{C,D\}$ correspondingly. Let $\textrm{sim}^{(k)}(A,
B)$ and $\textrm{sim}^{(k)}(C, D)$ denote the similarity scores
that bipartite Simrank computes for the node pairs $(A,B)$ and
$(C, D)$ after $k$ iterations. Then,
\begin{itemize}
\item[(i)] $\textrm{sim}^{(k)}(A, B) > \textrm{sim}^{(k)}(C,D),\
\forall \ k>0$, and \item[(ii)]$\lim_{k \rightarrow \infty} \
\textrm{sim}^{(k)}(A, B) = \lim_{k \rightarrow \infty}\
\textrm{sim}^{(k)}(C,D)$ if and only if $C_1 = C_2 = 1$, where
$C_1,\ C_2$ are the decay factors of the bipartite Simrank
equations.
\end{itemize}
\begin{proof}
Similar arguments as in Theorem \ref{theorem1}. \qed
\end{proof}
\label{theorem4}
\end{theorem}

\newpage
\section{Evidence-based Simrank similarity scores on complete bipartite graphs}
\label{sec:appendix2}

\begin{theorem}
Consider the complete bipartite graph $K_{2,2}$ with nodes sets
$V_1 = \{a, b\}$ and $V_2 = \{ A, B\}$. Let $\textrm{sim}^{(k)}
(A,B)$ denote the similarity between nodes $A, B$ that
evidence-bsaed bipartite Simrank computes after $k$ iterations and
let $C_1, C_2$ denote the Simrank decay factors. Then:

\begin{itemize}
\item[(i)]$\textrm{sim}^{(k)} (A,B) = \left( \frac{1}{2} +
\frac{1}{3}\right) \cdot \frac{C_2}{2}\sum_{i=1}^{k}
\frac{1}{2^{i-1}}C_1^{\lfloor\frac{i}{2}\rfloor}
C_2^{\lceil\frac{i-1}{2}\rceil}$

\item[(ii)] If $C_1, C_2 > \frac{1}{2}$ then $\lim_{k \rightarrow
\infty}\textrm{sim}^{(k)} (A,B) \geq \frac{C_2}{2}$
\end{itemize}

\begin{proof}

\begin{itemize}
\item[(i)] It follows directly from the definition of
evidence-based Simrank (Equations
\ref{eq:query_similarities_fixed} and
\ref{eq:ad_similarities_fixed}) and Theorem \ref{theorem1}

\item[(ii)] We have:
\begin{eqnarray}
\lim_{k \rightarrow \infty}\textrm{sim}^{(k)} (A,B) & = & \left(
\frac{1}{2} + \frac{1}{3}\right) \cdot \frac{C_2}{2} \lim_{k
\rightarrow \infty}\sum_{i=1}^{k}
\frac{1}{2^{i-1}}C_1^{\lfloor\frac{i}{2}\rfloor}
C_2^{\lceil\frac{i-1}{2}\rceil} \nonumber\\
 & > & 0.41666 \cdot C_2 \lim_{k
\rightarrow \infty}\sum_{i=1}^{k} \frac{1}{2^{2i}}\nonumber\\
& = & 0.41666  \cdot C_2 \cdot \frac{4}{3}\nonumber \\ & = &
0.5555 \cdot C_2
> \frac{C_2}{2}\nonumber
\end{eqnarray}
\qed
\end{itemize}
\end{proof}
\label{theoremb1}
\end{theorem}

\begin{theorem}
Consider the two complete bipartite graphs $G = K_{1,2}$ and $G' =
K_{2,2}$ with nodes sets $V_1 = \{a\}, V_2 = \{A, B\}$ and $V'_1 =
\{b, c\}$ and $V'_2 = \{C,D\}$ correspondingly. Let
$\textrm{sim}^{(k)}(A, B)$ and $\textrm{sim}^{(k)}(C, D)$ denote
the similarity scores that bipartite evidence-based Simrank
computes for the node pais $(A,B)$ and $(C, D)$ after $k$
iterations. Then, if for the
 decay factors of bipartite Simrank $C_1,\ C_2$ we know that $C_1, C_2 > \frac{1}{2}$
 we have

\begin{itemize}

\item[(i)] $\textrm{sim}^{(k)}(A, B) < \textrm{sim}^{(k)}(C,D),\
\forall \ k > 1$, and

\item[(ii)]   $\lim_{k \rightarrow \infty} \ \textrm{sim}^{(k)}(A,
B) < \lim_{k \rightarrow \infty}\ \textrm{sim}^{(k)}(C,D)$
\end{itemize}

\begin{proof}

\begin{itemize}
\item[(i)]
 It follows directly from the definition of
evidence-based Simrank (Equations
\ref{eq:query_similarities_fixed} and
\ref{eq:ad_similarities_fixed}) and Theorem \ref{theoremb1}
\item[(ii)] From Theorem \ref{theoremb1} we have:

\begin{equation*}
\lim_{k \rightarrow \infty}\textrm{sim}^{(k)} (C,D) \geq
\frac{C_2}{2}
\end{equation*}

Also, from the definition of evidence-based Simrank (Equations
\ref{eq:query_similarities_fixed} and
\ref{eq:ad_similarities_fixed}) we have:

\begin{equation*}
 \lim_{k
\rightarrow \infty}\textrm{sim}^{(k)} (A,B) = \frac{C_2}{2}
\end{equation*}

Thus, $\lim_{k \rightarrow \infty} \ \textrm{sim}^{(k)}(A, B) <
\lim_{k \rightarrow \infty}\ \textrm{sim}^{(k)}(C,D)$. \qed
\end{itemize}

\end{proof}
\label{theoremb2}
\end{theorem}

\begin{theorem}
Consider the two complete bipartite graphs $G = K_{m,2}$ and $G' =
K_{n,2}$ with $m < n$ and nodes sets $V_1, V_2 = \{A, B\}$ and
$V'_1, V'_2 = \{C,D\}$ correspondingly. Let $\textrm{sim}^{(k)}(A,
B)$ and $\textrm{sim}^{(k)}(C, D)$ denote the similarity scores
that bipartite evidence-based Simrank computes for the node pairs
$(A,B)$ and $(C, D)$ after $k$ iterations and let $C_1, C_2 >
\frac{1}{2}$, where $C_1,\ C_2$ are the decay factors of the
bipartite Simrank equations. Then,
\begin{itemize}
\item[(i)] $\textrm{sim}^{(k)}(A, B) < \textrm{sim}^{(k)}(C,D),\
\forall \ k>1$, and \item[(ii)]$\lim_{k \rightarrow \infty} \
\textrm{sim}^{(k)}(A, B) < \lim_{k \rightarrow \infty}\
\textrm{sim}^{(k)}(C,D)$
\end{itemize}
\begin{proof}
Similar arguments as in Theorem \ref{theoremb2}. \qed
\end{proof}
\label{theoremb3}
\end{theorem}

\end{document}